\documentclass[epsfig,preprint2]{aastex}

\usepackage{epsfig}

\newcommand \caII{\ion{Ca}{2}~}
\newcommand \feII{\ion{Fe}{2}~}
\newcommand \feIIn{\ion{Fe}{2}}
\newcommand \feI{\ion{Fe}{1}~}
\newcommand \heI{\ion{He}{1}~}
\newcommand \heIn{\ion{He}{1}}
\newcommand \liI{\ion{Li}{1}~}
\newcommand \liIn{\ion{Li}{1}}
\newcommand \hal{H$\alpha$~}
\newcommand \haln{H$\alpha$}
\newcommand \hbeta{H$\beta$~}
\newcommand \hbetan{H$\beta$}
\newcommand \hgam{H$\gamma$~}
\newcommand \hdel{H$\delta$~}
\newcommand \hdeln{H$\delta$}
\newcommand \msun{M$_\odot$}
\newcommand \kms{km s$^{-1}$~}
\newcommand \kmsn{km s$^{-1}$}

\shortauthors{Alencar \& Basri}  
\shorttitle{Profiles of Strong Permitted Lines in Classical T Tauri Stars}

\begin{document}

\title{Profiles of Strong Permitted Lines in Classical T Tauri Stars
\footnote{Based on observations obtained at Lick Observatory.} }

\author{Silvia H. P. Alencar\altaffilmark{2,3} and Gibor Basri\altaffilmark{2}}
\altaffiltext{2}{Astronomy Dept., Univ. of California, Berkeley, CA 94720-3411,
e-mail: silvia@crater.berkeley.edu, basri@soleil.berkeley.edu} 
\altaffiltext{3}{Depto. de F\' {\i}sica, ICEx, UFMG, C.P. 702,
Belo Horizonte, MG, Brazil, 30123-970}

\begin{abstract}

We present a spectral analysis of 30 T Tauri stars observed with the Hamilton
echelle spectrograph over more than a decade. One goal is to test magnetospheric 
accretion model predictions. Observational evidence previously published 
supporting the model, such as emission line asymmetry and a high frequency of 
redshifted absorption components, are considered. We also discuss the relation between 
different line forming regions and search for good accretion rate indicators. 

In this work we confirm several important points of the models, such as the 
correlation between accretion and outflow, broad emission components that are 
mostly central or slightly blueshifted and only the occasional presence of 
redshifted absorption.  We also show, however,
that the broad emission components supposedly formed in the magnetospheric
accretion flow only partially support the models. Unlike the predictions, they
are sometimes redshifted, and are mostly found to be symmetric. 
The published theoretical profiles do not have a strong resemblance to our observed ones.
We emphasize the need for accretion models to include a strong turbulent component 
before their profiles will match the observations. The effects of rotation, and 
the outflow components, will also be needed to complete the picture. 

\end{abstract}

\keywords{line: profiles --- stars: pre-main-sequence --- stars: formation}

\section{Introduction}

Classical T Tauri stars (CTTSs) are young, almost solar-mass stars that exhibit
a wide range of permitted and sometimes also forbidden emission lines, together
with an excess continuum emission that goes from the infrared to the ultraviolet.
Their spectral energy distribution is consistent with the presence of a circumstellar 
disk that appears to play a major role in the regulation of both the infall
and outflow of material from the star-disk system.
A strong wind component is also thought to be present and cause the blueshifted
absorption features commonly seen in the Balmer lines. 

In the past years many attempts have been made to explain those characteristics. 
The permitted line emissions were discussed in terms of outflowing winds \citep{hart82, 
hart90,natta90}, turbulence in a boundary layer 
between the disk and the stellar surface \citep{bertout88,basri89} 
and chromospheric activity \citep{calvet84,calvet85}. More recently, 
magnetospheric accretion models were proposed \citep{hart94,shu94,hart98}. 
In these models the circumstellar disk is truncated by stellar 
magnetic field lines and the strong broad emissions arise from the accelerated
infalling material channeled through the lines that connect the disk to the star.
At the base of the magnetospheric accretion column, where the accreted material 
hits the star, hot spots and the hot continuum emission (veiling) are produced.

\citet{ehg94} showed observational evidence in a sample of 15 CTTSs
that they claimed confirm some of the magnetospheric predictions, 
like blueward asymmetric emission 
lines (due to disk occultation of the receding flow) and redshifted absorption 
components at typical free-fall velocities (inverse P Cygni profiles), which are naturally 
explained by the magnetospheric infall.
The line profiles of several stars were also well reproduced by magnetospheric
calculations \citep{muze,najita}, but often the theoretical profiles 
are more asymmetric and always less broadened than the observed ones. 
However, we must keep in mind that the models of \citet{hart94}
do not include rotation and winds that would certainly influence both the asymmetry and 
the broadening of the theoretical profiles.

The CTTSs were initially found to be slow rotators due to observational surveys
that reported a bimodal distribution of stellar rotation periods among TTS
\citep{choi}, where
the slow rotators (P $>$ 4 days) seemed to exhibit disk signatures, as near
infrared excess emission, that the fast rotators did not \citep{edwards93}. 
It has always been a 
challenge to understand the spin-down of those stars but the magnetospheric model 
predicts that the magnetic interaction between the star and the disk could regulate the 
stellar rotation. This could explain why weak T Tauri stars (WTTS), that do not have
disks present a wider range of rotational periods. Recently \citet{stassun},
analyzing the rotation period distribution of young stars in the Orion nebula region,
showed that this bimodal distribution does not seem to exist and that the earlier 
works conclusion was also statistically consistent with a uniform distribution. 
\citet{herbst} suggest this recent result, however,  may be confined to low mass TTSs.

In view of the many points still under discussion, we present 
in this paper the spectral analysis of an observed sample of 30 TTSs with a large 
spectral coverage that are used to analyse the magnetospheric model's predictions
and test the previous observational results. 
The observations and reduction procedures are described in Section 2, the equivalent
width and veiling measurements are detailed in Section 3, we analyse the 
many features in our spectra in Section 4, discuss the results in Section 5 and draw 
our final conclusions in Section 6.  

\section{Observations }

We present the spectral analysis of a sample of 30 TTSs listed in Table \ref{eqw}.
The observations, which span over more than a decade, were carried out at 
Lick Observatory with the 3m Shane telescope and the Hamilton Echelle 
Spectrograph \citep{vog87} coupled either to a TI 800x800 CCD or a FORD 
2048x2048 CCD. 
With the smaller detector two settings were established: a red setting 
covering 52 orders from 4900\AA ~ up to 8900\AA ~ and a blue 
setting covering 38 orders from 3900 \AA ~ up to 5200\AA. 
The spectral coverage, however, is not complete, with a gap of a few 
dozens of angstroms between orders. Whenever possible, blue and red
observations were obtained in the same night or within
1 night of difference, but for some stars we only have the red
setting. The bigger CCD installed 
in 1992 permits a full spectral coverage of $\sim$ 92 orders ranging from 
3900 \AA ~ to 8900 \AA.
The mean resolution of the spectrograph is $\lambda / \Delta\lambda \approx
48,000$ and the exposure times varied from 15min to 1h15min, depending on the
target and on the CCD used. 

The reduction was performed in a standard way described by \citet{val94}
which includes flatfielding with an incandescent lamp exposure,
background subtraction and removal of cosmic rays. Wavelength calibration
is made by observing a Thorium-Argon comparison lamp and performing 
a 2-D solution to the Thorium lines. Radial and barycentric velocity
corrections are applied and all the data shown are in the stellar rest frame. 
The spectra are not flux calibrated, so each spectrum has been continuum 
normalized. Due to differences in weather conditions, exposure times
and efficiency between different chips there is a wide range of signal to
noise in the data. However, not only strong emission line profiles were 
reliably extracted but also many absorption line profiles as shown in Figure
\ref{profiles}. 
After the reduction procedure the spectra have been binned and smoothed
with a window width of 3.
A weak-line T Tauri star was included in the sample, V410 Tau \citep{hbc}, 
as a reference of stars without disk accretion.
DQ Tau is a known binary system \citep{bjm}, the chosen spectrum 
corresponds to an outburst event when the lines are stronger and to a phase 
($\phi=0.02$) where there is no separation between the lines of the two 
components. 

\begin{figure*}[htb]
\begin{center}
\epsfig{file=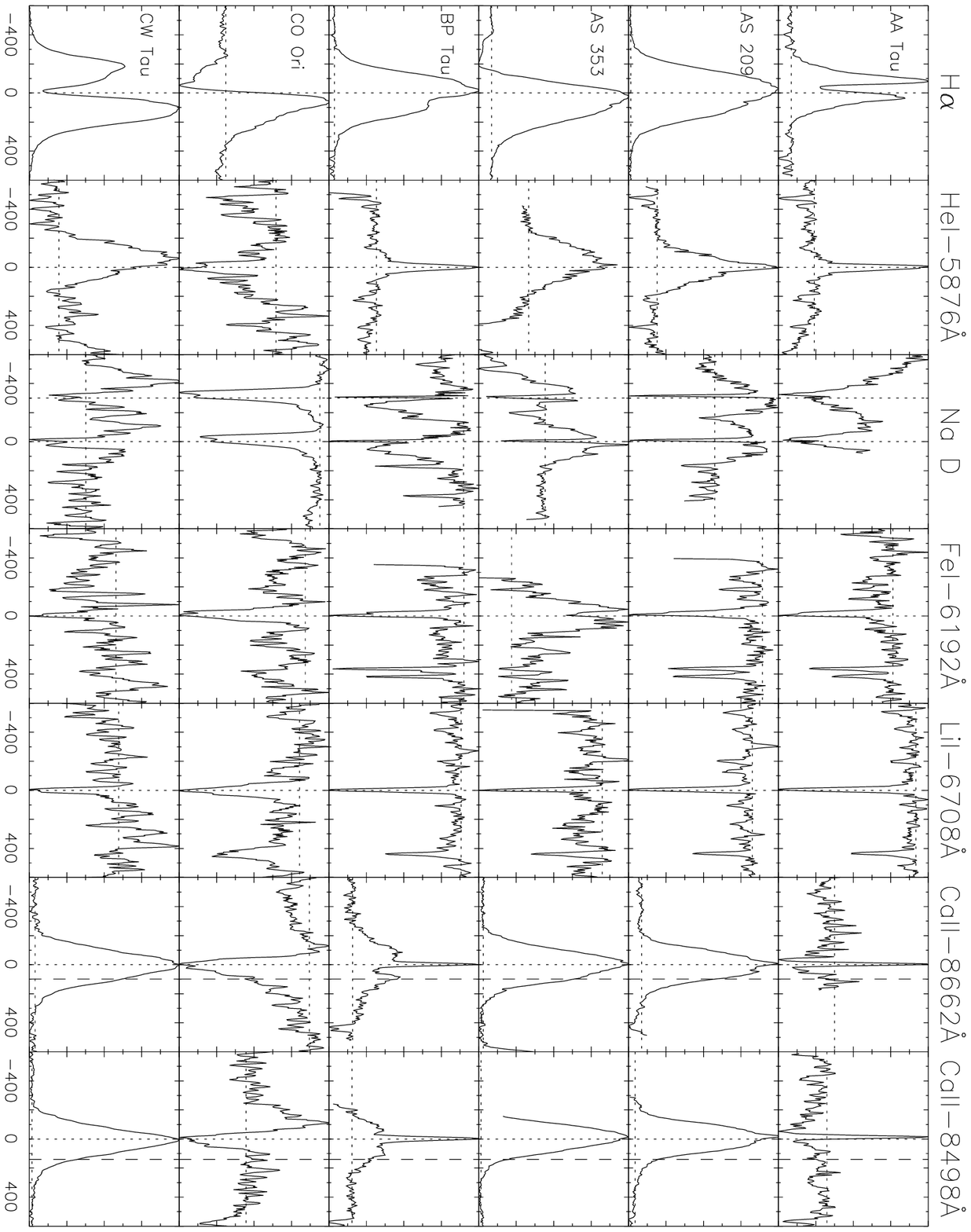,height=20cm}
\vspace{-1cm}
\caption{\label{profiles} Line profiles - all the data have been continuum
normalized, binned and smoothed with a window width of 3. The horizontal
dotted lines represent the continuum level at 1. The vertical dotted lines
are the spectral line's center at the stellar rest frame. The
contamination of the Na D lines by the vapor lamps of the city of San Jose
has been removed. The dashed lines close to the IRT profiles are the
velocity position of the blended Paschen emission lines.}
\end{center}
\end{figure*}

\setcounter{figure}{0}
\begin{figure*}[htb]
\begin{center}
\epsfig{file=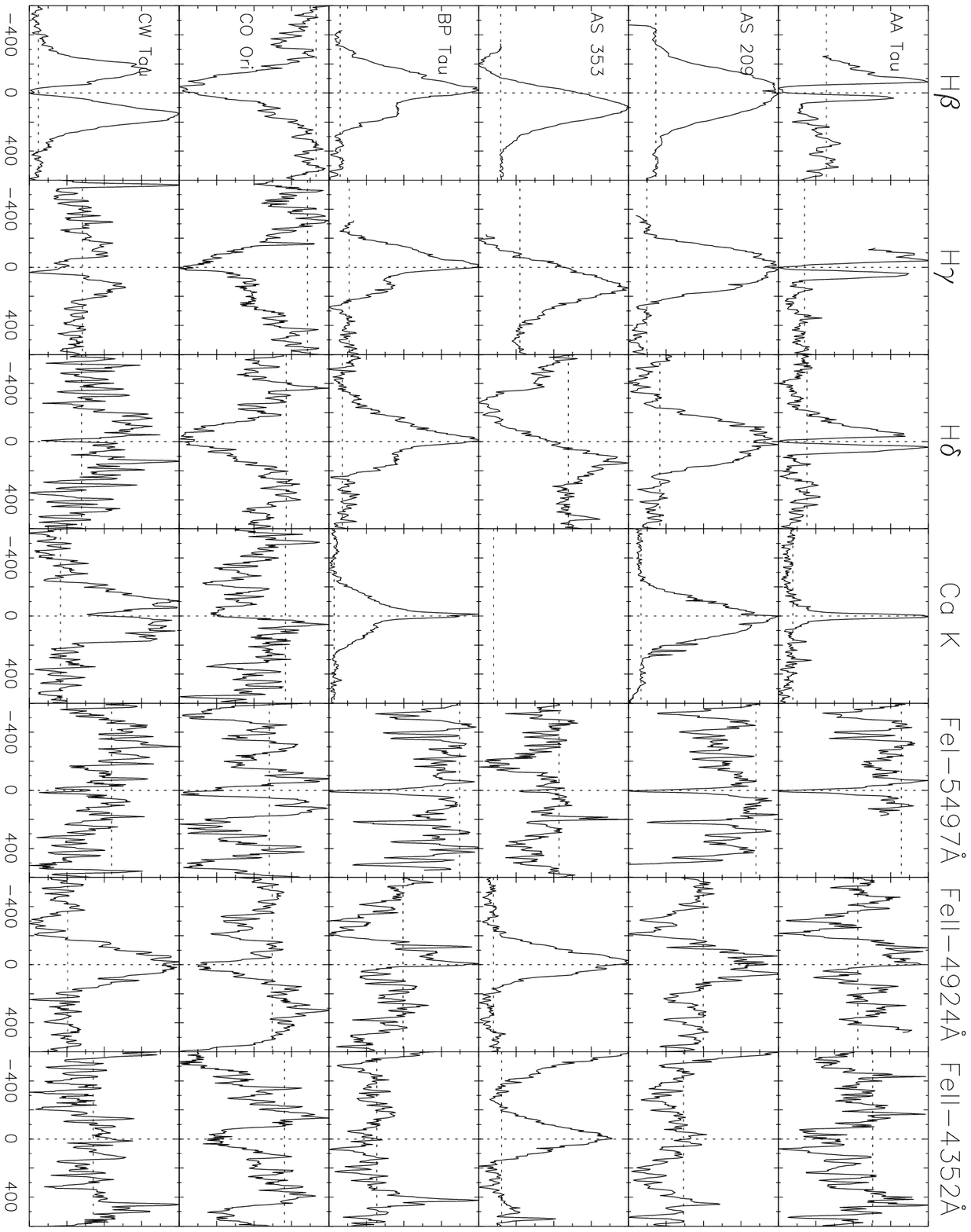,height=20cm}
\caption{Continued}
\end{center}
\end{figure*}

\setcounter{figure}{0}
\begin{figure*}[htb]
\begin{center}
\epsfig{file=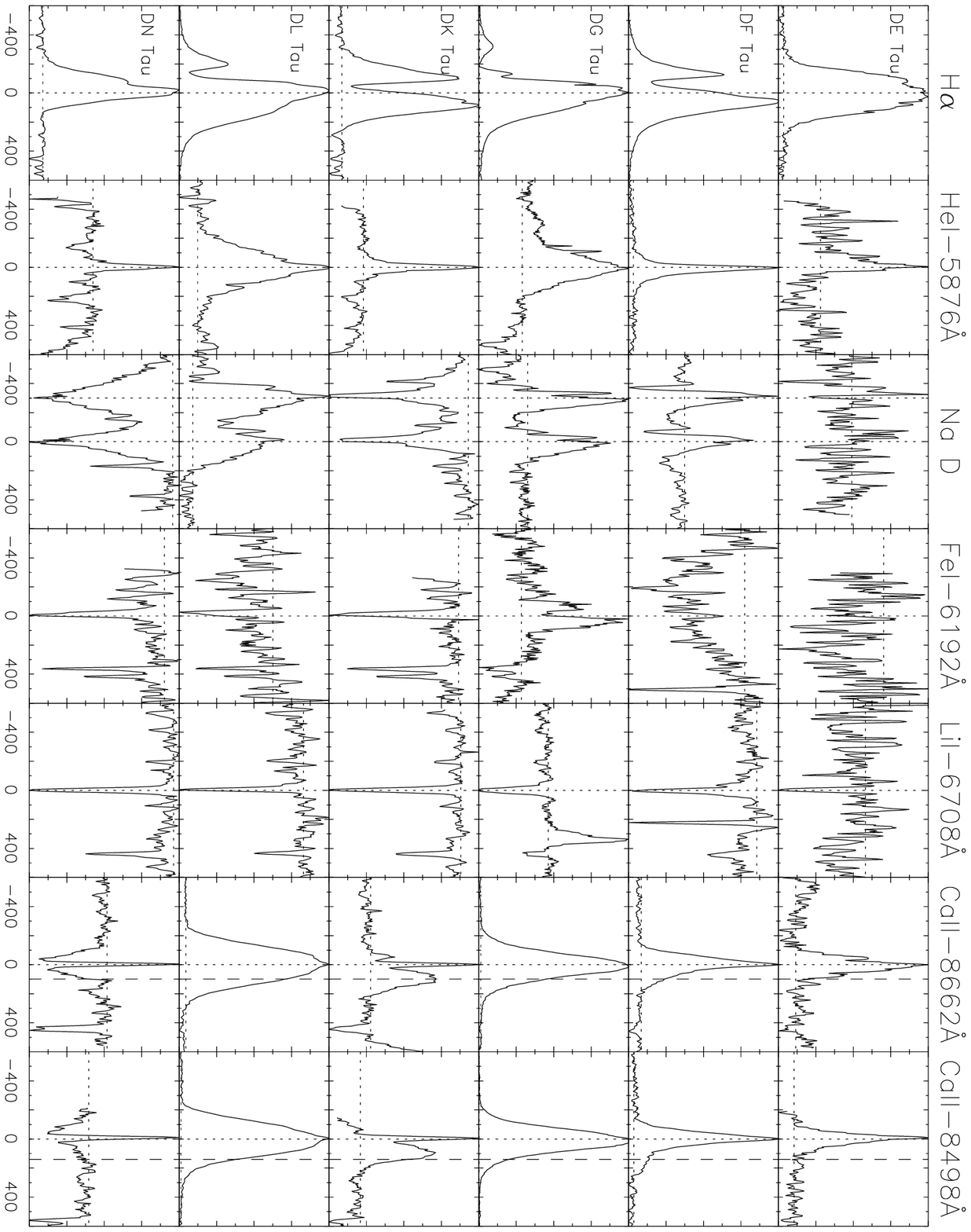,height=20cm}
\caption{Continued}
\end{center}
\end{figure*}

\setcounter{figure}{0}
\begin{figure*}[htb]
\begin{center}
\epsfig{file=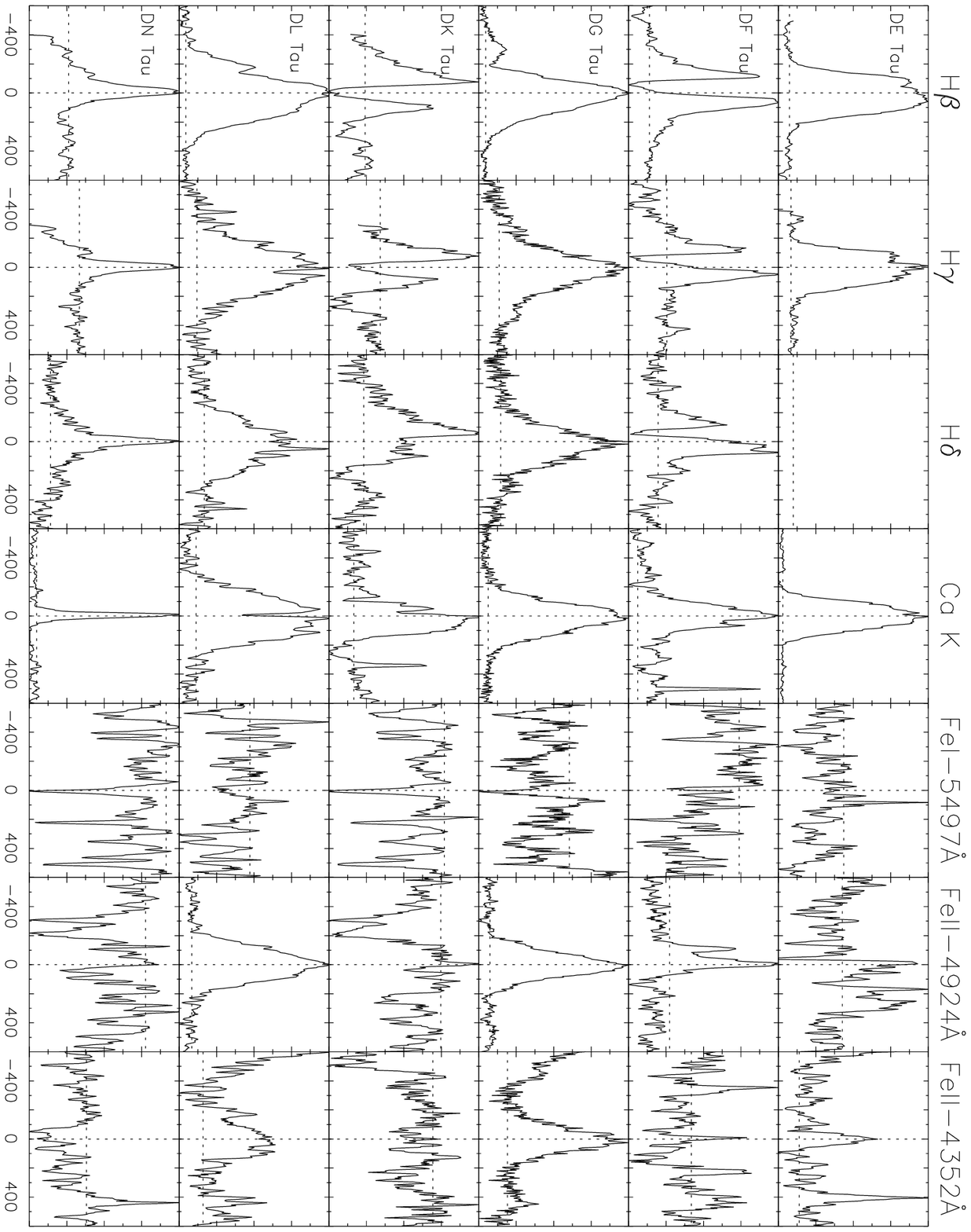,height=20cm}
\caption{Continued}
\end{center}
\end{figure*}

\setcounter{figure}{0}
\begin{figure*}[htb]
\begin{center}
\epsfig{file=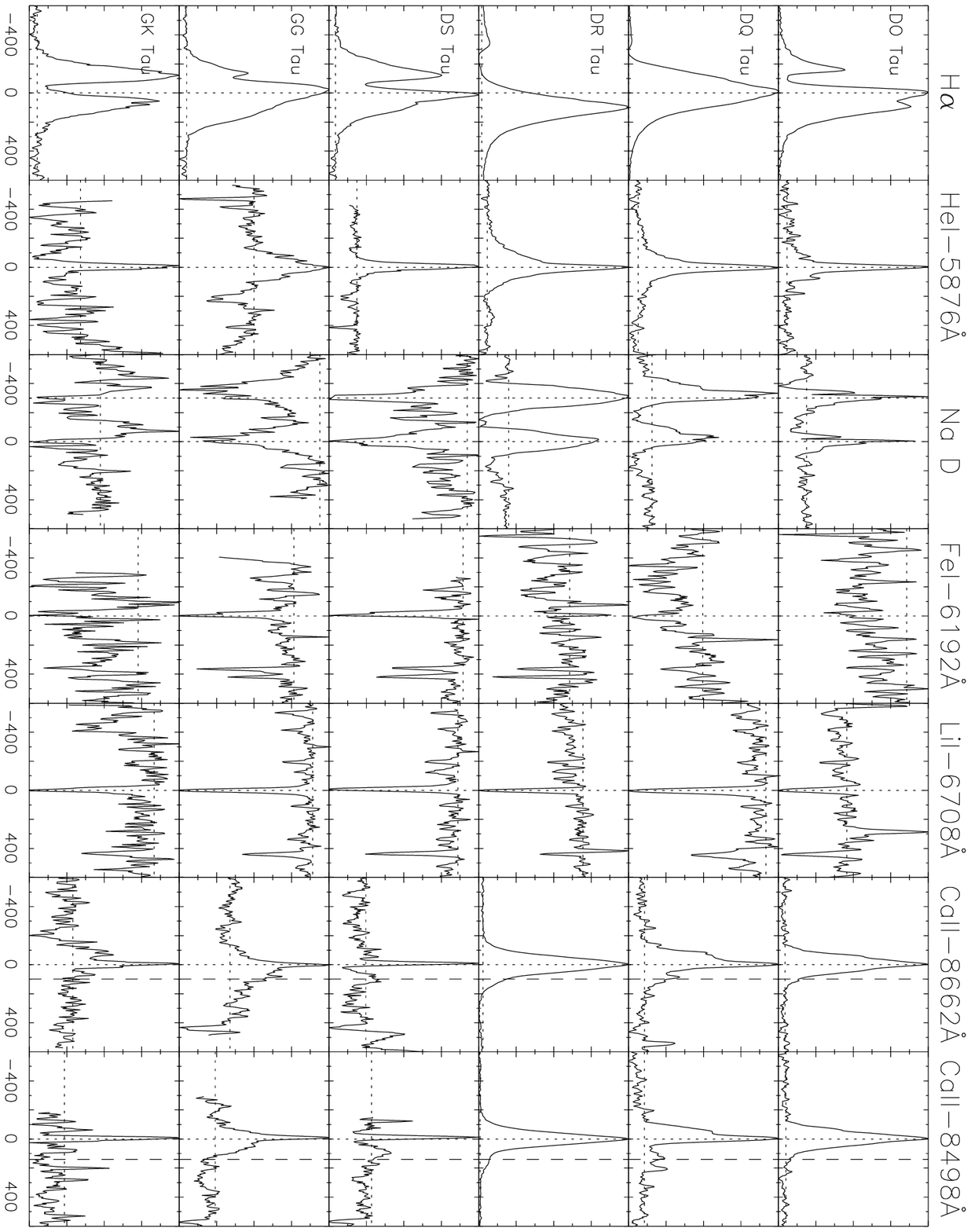,height=20cm}
\caption{Continued}
\end{center}
\end{figure*}

\setcounter{figure}{0}
\begin{figure*}[htb]
\begin{center}
\epsfig{file=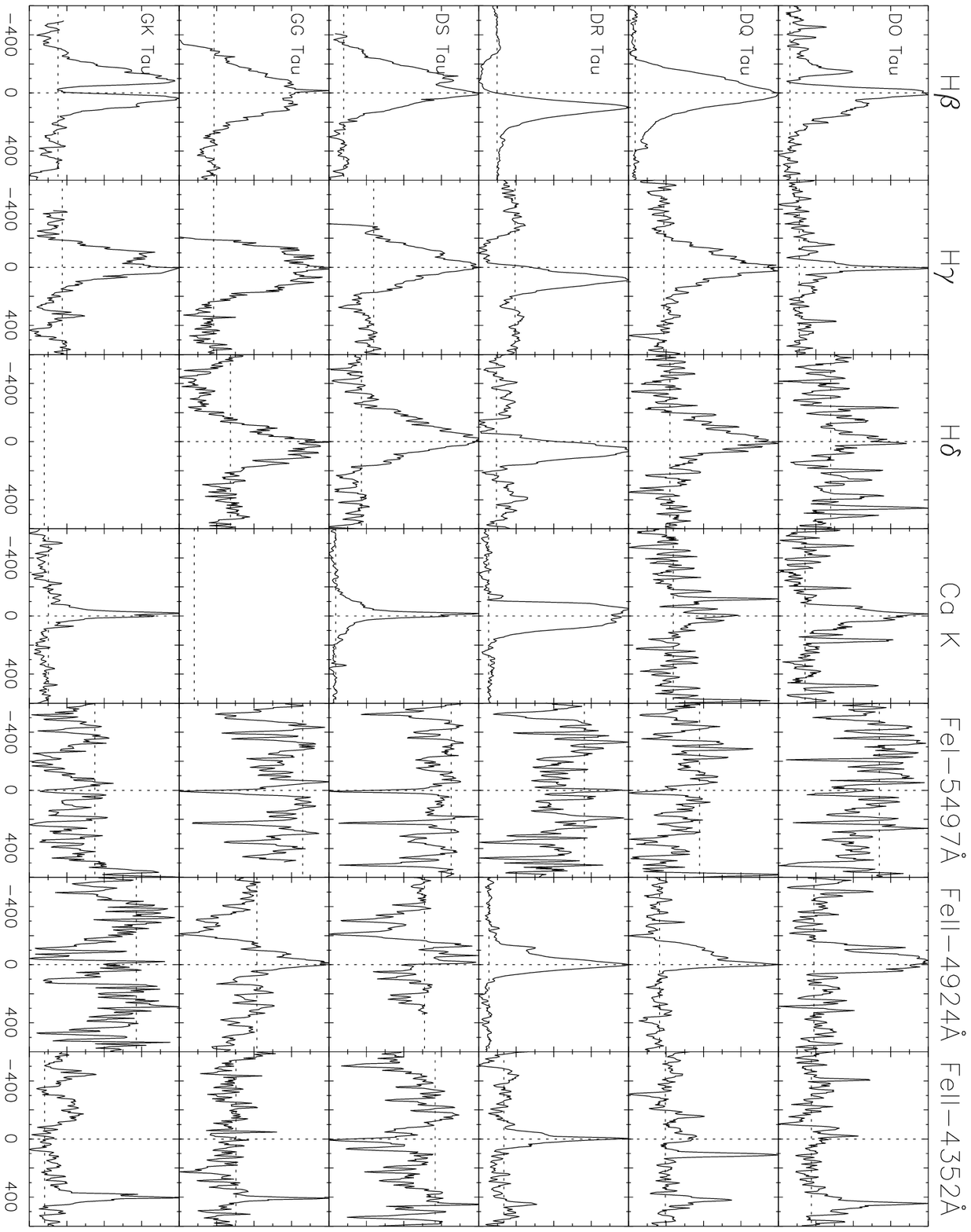,height=20cm}
\caption{Continued}
\end{center}
\end{figure*}

\setcounter{figure}{0}
\begin{figure*}[htb]
\begin{center}
\epsfig{file=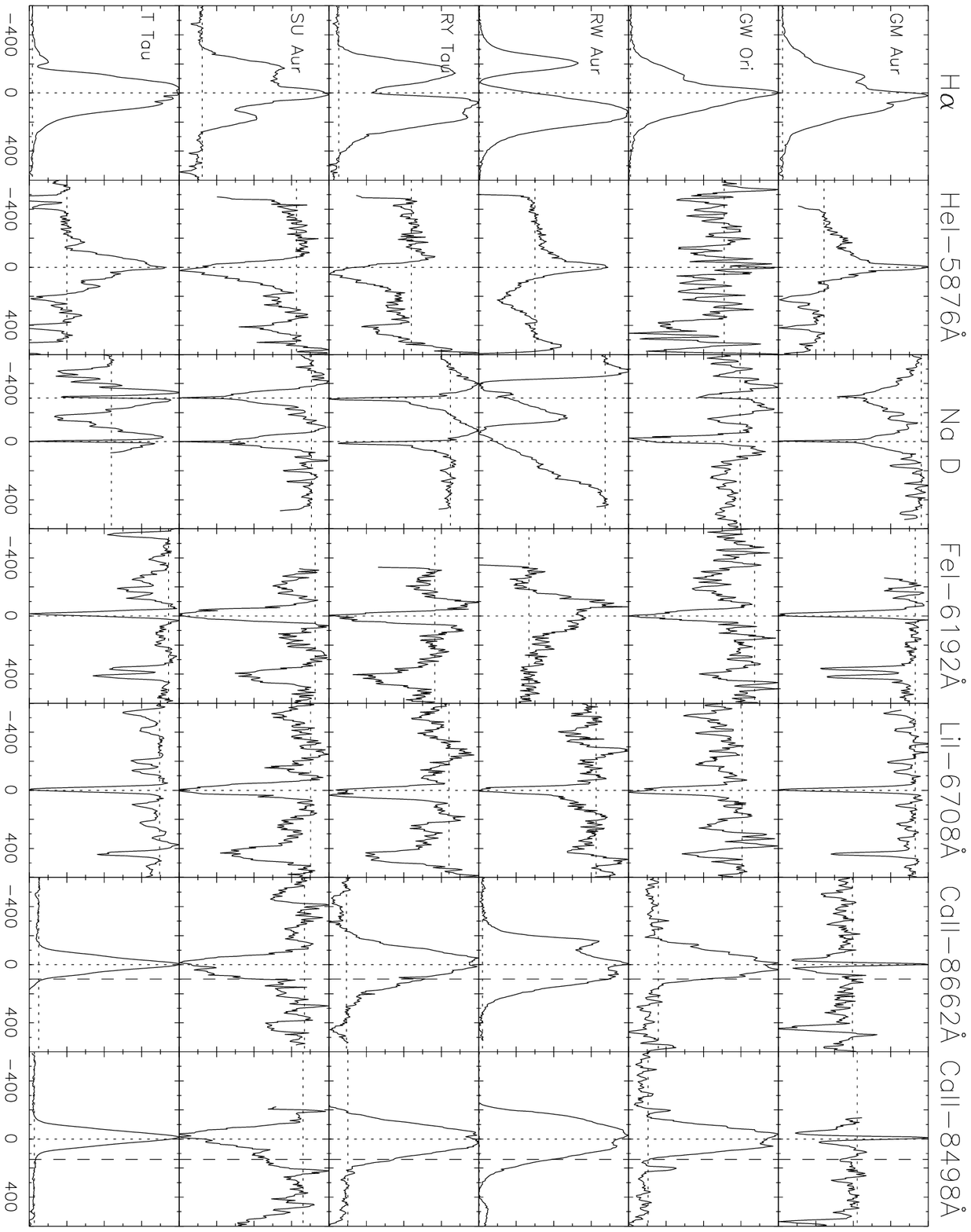,height=20cm}
\caption{Continued}
\end{center}
\end{figure*}

\setcounter{figure}{0}
\begin{figure*}[htb]
\begin{center}
\epsfig{file=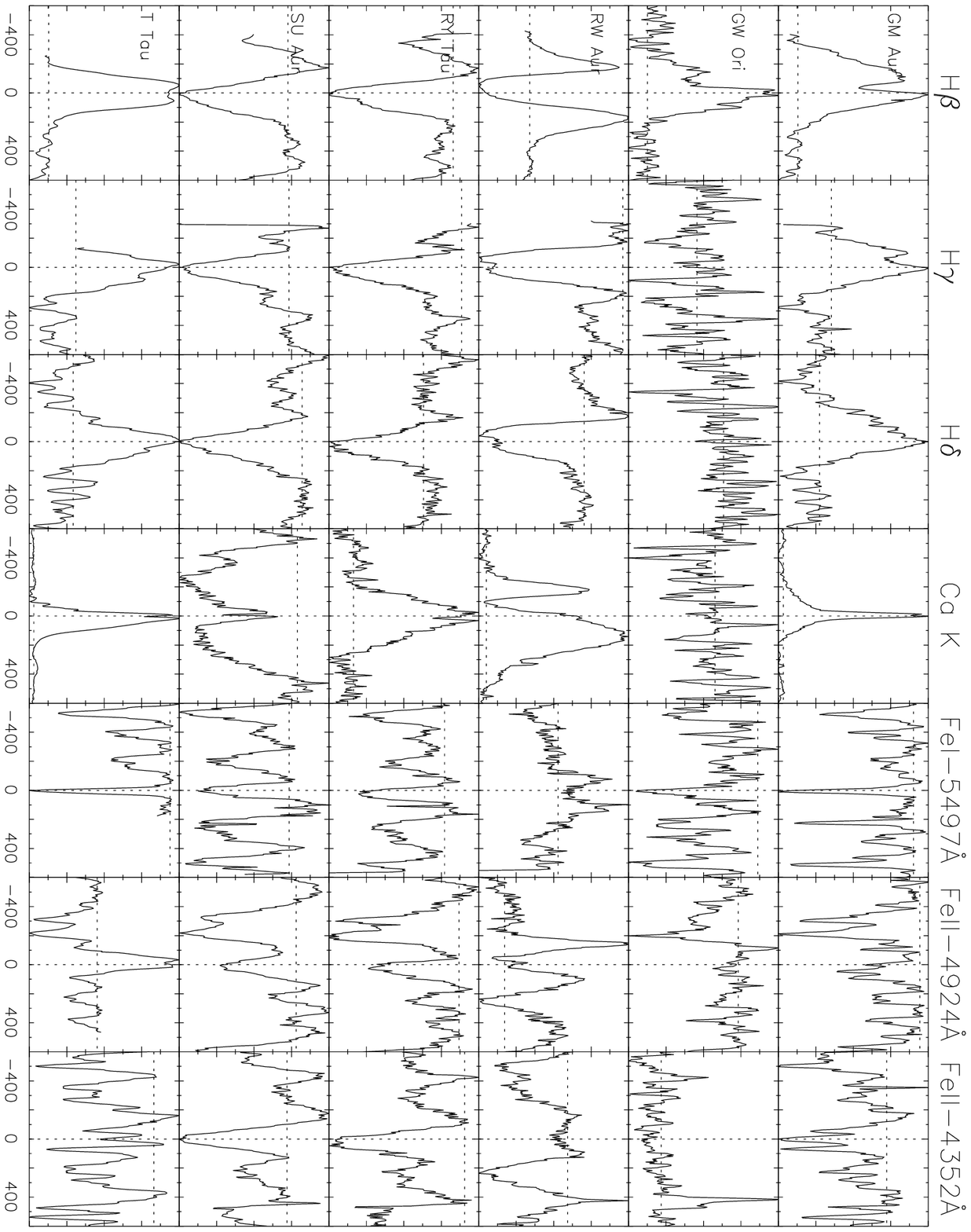,height=20cm}
\caption{Continued}
\end{center}
\end{figure*}

\setcounter{figure}{0}
\begin{figure*}[htb]
\begin{center}
\epsfig{file=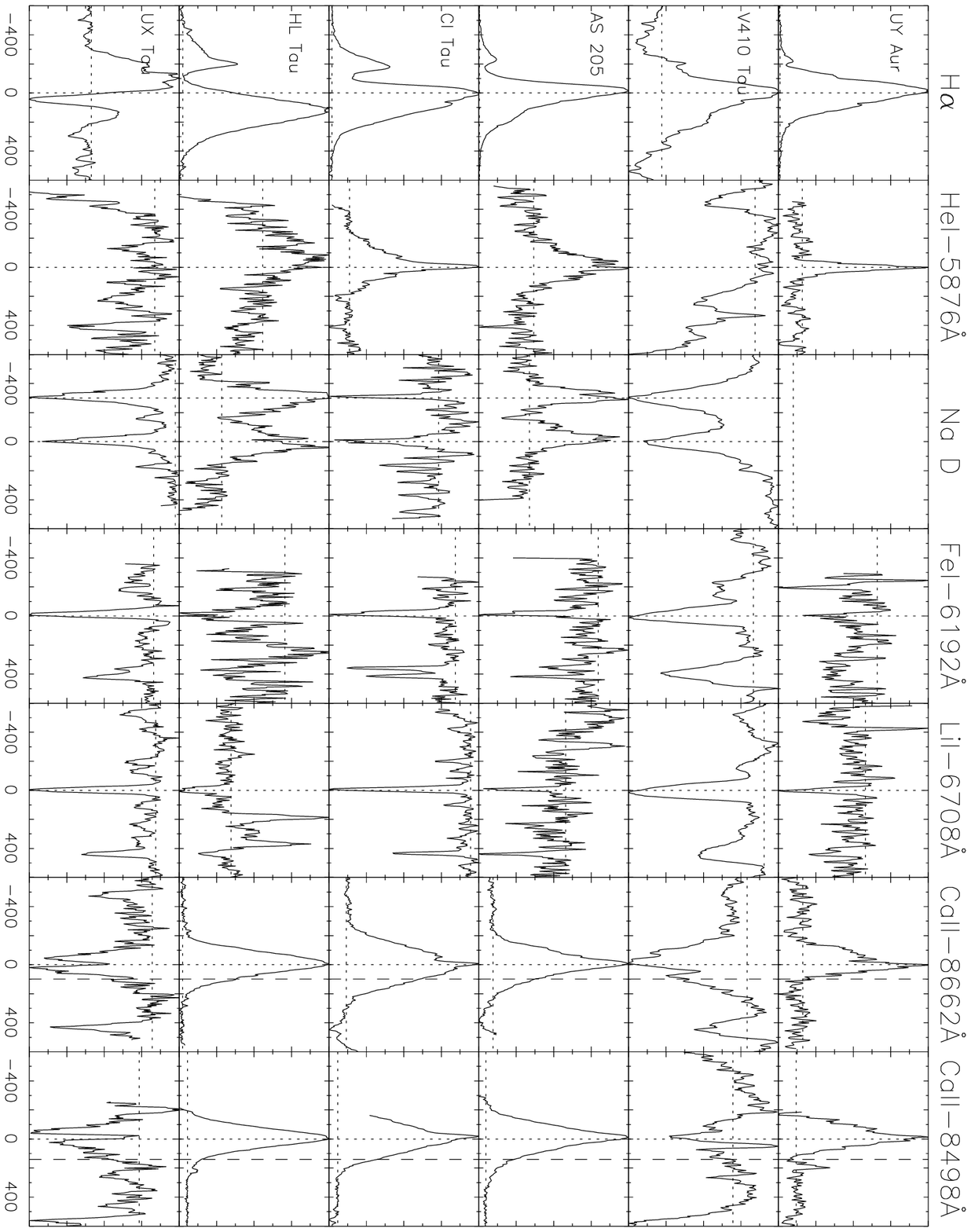,height=20cm}
\caption{Continued}
\end{center}
\end{figure*}

\setcounter{figure}{0}
\begin{figure*}[htb]
\begin{center}
\epsfig{file=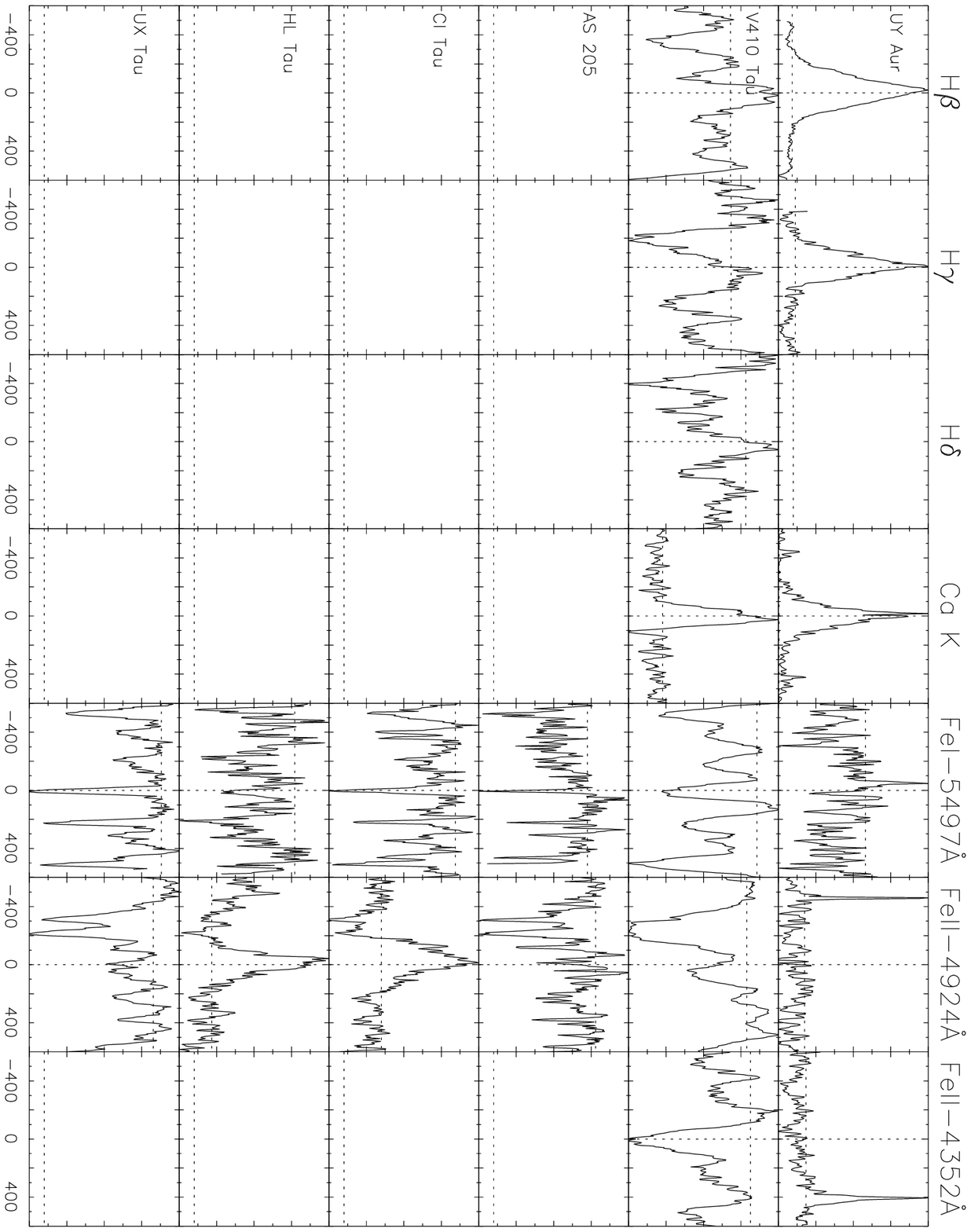,height=20cm}
\caption{Continued}
\end{center}
\end{figure*}

\section{Measurements}

\subsection{Equivalent widths and profile decomposition}
We first measure the equivalent width of all
the lines in Figure \ref{profiles} that could be reliably identified. The results
are presented in Table \ref{eqw} and will be used in the following sections to
study the relations between ``fluxes'' (veiling corrected equivalent
widths) from different lines and between fluxes and parameters such as
the mass accretion rate.

The line profiles were also decomposed %into components
with a multiple gaussian procedure based on the Marquardt method,
so that we could also study correlations between 
line components. 
Most of the lines were very well fitted by 1 or 2 gaussians
and only sometimes required a third component. Some exceptions are the 
flat topped profiles of T Tau, SU Aur and DE Tau that are clearly non-gaussians.
The fitting procedure was not always straightforward, since some lines are contaminated 
by non-stellar emission, some are blended by other lines
and some can be fitted with more than one gaussian combination (like 2 emissions
instead of 1 emission and 1 absorption, for example). We explain below how we 
dealt with those problems.

The \hal line often exhibits a double peaked profile which was, in most cases, better
fitted by a broad emission and a blue absorption than 2 emissions. The \hal 
wings are not steep and seem to belong to a very broad emission line rather than to 2 
narrow ones. It is often possible to confirm that a true absorption is
present by looking at the upper Balmer lines where the feature is enhanced 
relatively to the emission and dips below the continuum. 

The Na D lines are strongly contaminated by the vapor lamps of the city of
San Jose but our decomposition procedure was able to reliably subtract
that contamination with a simultaneous gaussian fit 
while decomposing the line profiles. 

The \caII and \heI lines exhibit profiles typically with narrow and 
broad emission components; some lines show only one of the components. In some
\caII lines the narrow component is present inside a photospheric 
absorption core and, in that case, the emission-line equivalent width was 
measured from the base of the absorption core instead of the continuum. 
The \caII infrared triplet (IRT) lines are also blended with a 
Paschen emission line that has usually strengths less than 20\% of the IRT, 
as also noted by \citet{hp}. Whenever the \caII lines are weak or narrow, 
the Paschen contamination is significant and has been extracted from the 
equivalent width measurements. 

Finally, the \feII equivalent widths were only measured when the line was in
strong emission and could surely be distinguished from the lines nearby. 

\subsection{Veiling}
The equivalent widths presented in Table \ref{eqw} are contaminated by the variable 
continuum against which they were measured. They need to be corrected for
the veiling in order to represent a true emission strength that can be
regarded as a ``flux'' measurement.
Some of our spectra have measured veilings in previous papers \citep{bb90,
jb97} but in order to be consistent we decided to measure all the
veilings again. The new values are presented in the last columns of Table \ref{eqw} 
and the common measurements agree within the errors with the previously published 
ones.

In order to calculate the veilings we chose spectral orders with many 
photospheric absorption lines both in the blue and red settings.
The red orders correspond to the wavelength ranges 5557\AA $< \lambda <$ 5642\AA~ and
5668\AA $< \lambda <$ 5755\AA~ and the blue orders to
4571\AA $< \lambda <$ 4641\AA~ and 4763\AA $< \lambda <$ 4836\AA.
The idea is to compare the spectrum of a CTTS with that
of a standard star with the same spectral type that has been
broadened to the CTTS rotational velocity and veiled.
The choice of suitable standard stars is important and we
followed, whenever possible, the suggestion of \citet{bb90} to use the 
Hyades dwarfs as reliable standards. We also used the standard and CTTS
spectral types determined by them.

We first rotationally broadened the continuum normalized spectrum of the 
standard star and then applied the veiling to the standard's spectrum, 
comparing it with the observed CTTS spectrum until a good match was found. 
The best adjustment was determined by eye, but a unique veiling value did 
not perfectly match all the photospheric lines and the results presented
in Table \ref{eqw} correspond to a mean value obtained with all the lines in the 
selected orders.   
The errors in the veiling determination are bigger for the higher veiling 
values, as the photospheric lines of highly veiled stars are extremely 
shallow and hard to match. Typically
we have errors of 0.1 for veilings smaller than 1.0 and up to 0.5 for the
higher ones.
Most of the rotational velocities were taken from \citet{bb90}, although
we verified them by checking the match between the FWHM of
the cross-correlation normalized profiles of the original standard with the CTTS 
spectra and of the original standard with the broadened and veiled standard. 

The veiling values obtained are used to search
for correlations between veiling corrected line equivalent widths, defined
as in \citet{jb97}:

\begin{equation}
W^0_{\rm eq}=W_{\rm eq}(V+1)
\end{equation}
where $W_{\rm eq}$ is the measured equivalent width and $V$ is the veiling.

\section{Analysis}

\subsection{General Characteristics}
The magnetospheric accretion model is the current consensus model
to describe the accretion processes in CTTSs. These models predict general 
trends like central or blueshifted asymmetric broad emission lines, sometimes 
with redshifted absorption components \citep{hart94,muzeb}, 
that we should be able to verify in our large sample of stars and lines.
The magnetospheric accretion scenario may also include winds that would cause 
further emission and blueshifted absorption \citep{shu94}.
Some theoretical predictions have been confirmed by observational results \citep{ehg94,
muze} but there is still a long way to go in order to fully understand
the nature of the processes taking place in CTTSs.
Below are some general profile characteristics from 
our sample of stars that we will analyse in detail in the next sections.  

Many of our emission lines show broad components (BC) that tend to have
blueshifted centroids (\haln, \hbetan, the \heI BC, the IRT BC, the Na D and the 
\feII $\lambda$4923) following the magnetospheric
predictions, but the centroids are also 
redshifted in some stars and exhibit a wide range of shift values. The broad
emission components are also generally well fitted by gaussians which may have a turbulent 
origin \citep{basri90} rather than the infall scenario. 
Redshifted absorption is not commonly present. The clearest example is 
BP Tau, where it appears in the Balmer lines, the Na D and the \caII ($\lambda$8662). 
Blueshifted absorption is a much more common feature that can be seen in many 
Balmer and Na D lines and sometimes in lines like Ca K in RW Aur. 
Many lines show a multiple-component profile that can often be decomposed 
either as narrow and broad components, e.g. \heIn, Ca 
and Fe lines \citep{bb90,bek}, or as emission and absorption components, 
like the Balmer lines. The different components often show distinct characteristics
(like broadening and shifts) suggesting that the lines may originate in distinct
regions.
The \feI lines are mostly in absorption but can be found in strong emission in very
active stars
(as in AS 353, DG Tau and RW Aur) always with a double-peaked profile 
that could be due to NLTE chromospheric core inversion (inversion of the source function)
or disk rotation.

\subsection{Line Strengths}

The study of the correlations between the fluxes of different lines is an
important tool that gives insight on the relations between different line formation
regions. Correlations can indicate common forming regions but can also mean that
a common process is affecting both regions at the same time.
In this section we discuss the correlations that
appear and also those that do not in our data.

\begin{figure}[htb]
\begin{center}
\epsfig{file=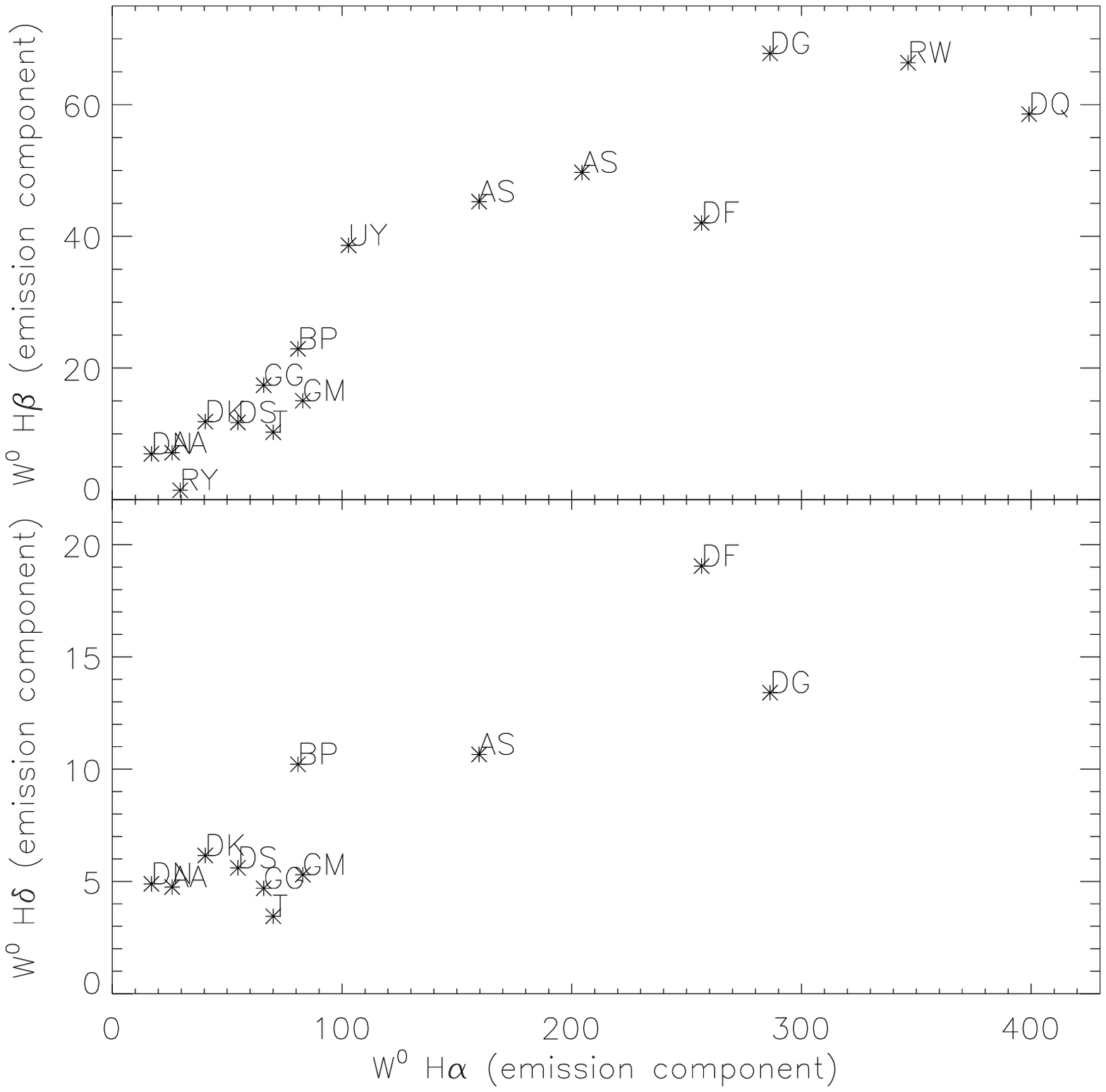,width=8cm}
\caption{\label{wbalmer} Veiling corrected equivalent widths of the emission components:
$W^0_{\rm H\alpha}$ vs. $W^0_{\rm H\beta}$ and vs. $W^0_{\rm H\delta}$.}
\end{center}
\end{figure}

The Balmer lines emission components are well correlated with each other, as are 
the IRT lines with each other, which is expected as
they should be formed in regions with the same physical characteristics. 
The difference between them 
should arise mainly due to differences in line optical depths. 
In the Balmer lines case,
the correlation seems to decrease as we compare lines that are further apart
in the series, e.g. \hal and \hbeta show a strong correlation
with each other (Figure \ref{wbalmer}, top) with a false alarm probability (FAP) 
of $3.24\times10^{-7}$, but
\hal and \hdel do not (FAP=0.168). %use VnHdelta_j_12a, r=0.390
This behavior is biased by stars with strong wind
contribution, where the emission component is sometimes almost absent due to 
large absorption components and/or a hot underlying photosphere.
Eliminating those 
stars the good correlations are partially recovered (Figure \ref{wbalmer}, bottom), 
the FAP of the \hal and
\hdel emission components decreasing to $3.55\times 10^{-4}$ and presenting a 
linear correlation coefficient r=0.880.

\begin{figure}[htb]
\begin{center}
\epsfig{file=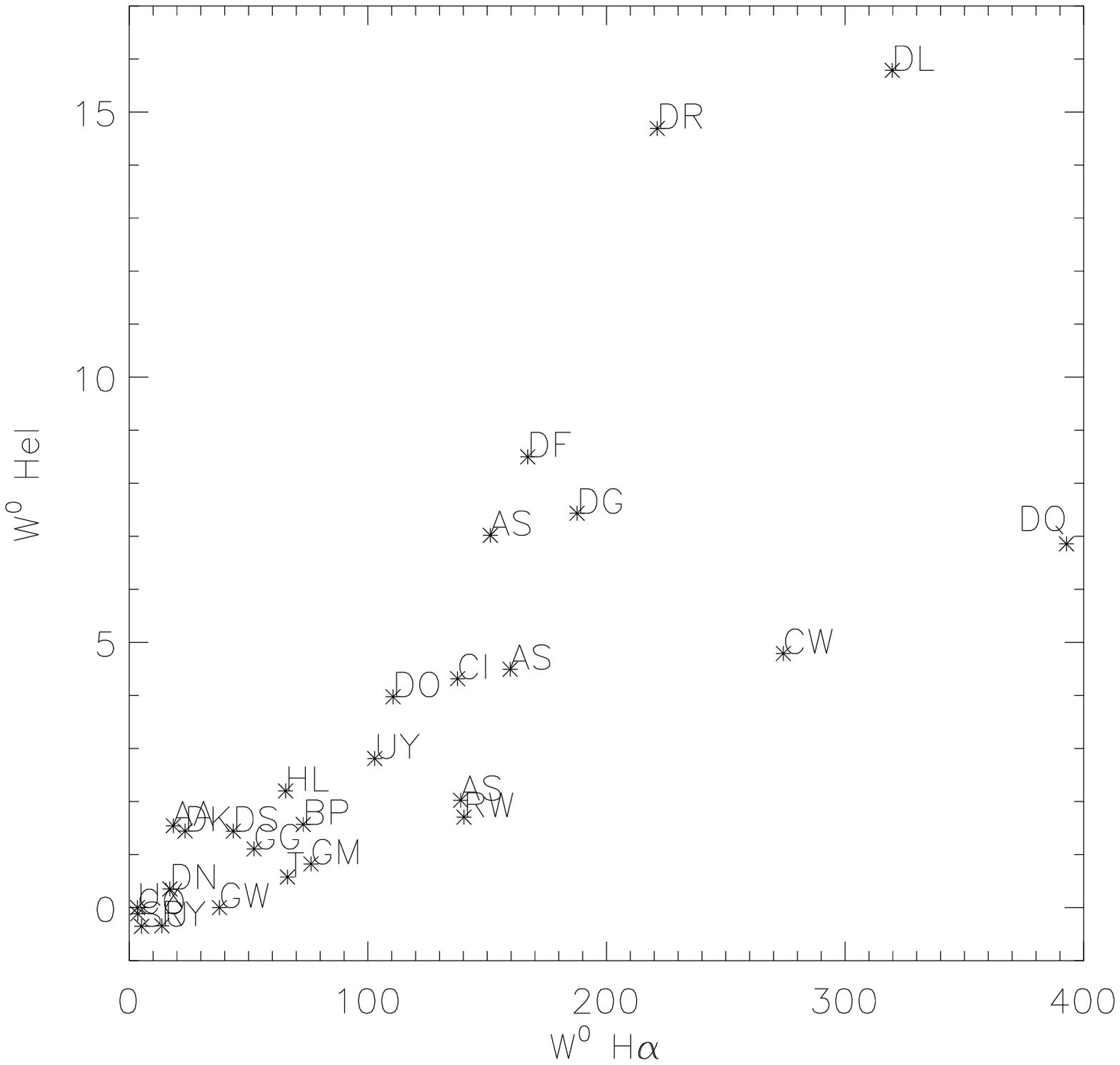,width=8cm}
\caption{\label{whalheI} $W^0_{\rm H\alpha}$ vs. $W^0_{\rm He I}$}
\end{center}
\end{figure}

The entire \hal and \heI lines are also well correlated (Figure \ref{whalheI} ; 
FAP=$1.26\times 10^{-6}$), %and r=0.786 
except for DQ Tau which has a very high \hal equivalent width in the outburst 
spectra we chose and CW Tau which has a high uncertainty in the veiling 
measurement (we get FAP=$4.65\times 10^{-10}$ without DQ Tau and CW Tau). %and r=0.906
The same kind of correlation is verified with \hbetan,
but \hgam and \hdel do not show a well defined correlation with \heIn.
The apparent lack of correlation with \hgam and \hdel may be due to
the small number of spectra where those lines could be reliably measured and
the lower S/N present in our spectra in that region.
Apart from the correlation with the entire lines, we also verified some good 
correlation between the \hal emission component and 
the \heI BC (FAP=$4.68\times 10^{-4}$ with DQ Tau and CW Tau and 
FAP=$8.32\times 10^{-6}$ without them). 
The \heI line is believed to form in high temperature regions (20,000K to 50,000K)
and together with the large width of its BC it suggests that the BC comes either
from the highly turbulent shock region or the magnetospheric 
flow, where the broad emission component of the Balmer lines is also supposed
to be formed. Due to their larger optical depths, the Balmer lines 
probe a much larger volume of gas than the \heI lines, but the correlations between
the entire lines and the components
show that their formation regions are strongly related to each other.

Also thought to be formed in the accretion
flow because of its large widths and blueshifted center, the broad IRT component 
does not strongly correlate with the \heI BC 
(FAP=$1.97\times 10^{-3}$ and r=0.678 for $\lambda$8498) 
or with the \hal emission (FAP=$6.12\times 10^{-3}$ and r=0.590 for $\lambda$8498). 
For the \heI BC, no correlation is found for small equivalent width
values and a tendency for the equivalent widths to increase together exists,
although with much scatter. For the \haln, however, a correlation seems to exist
except for the higher \hal flux values (where the scatter is really large).
Although the \caII line is formed at much lower temperatures, it is
expected to strongly correlate with 
\hal and \heI if the accretion flow is sufficiently organized that a global
change in accretion rate is felt by all diagnostics.

\citet{bsb96} showed that the narrow components (NC) of CTTSs
have a flux excess when compared to the WTTSs and
this was attributed to the reprocessing of radiation produced in accretion shocks
as the accreting material hits the stellar atmosphere.
However, neither the \heI NC nor the \caII NC correlate with the Balmer emission, 
which implies that 
although both the NCs and the Balmer lines are related to accretion, they 
are probably probing regions that are far away from each other and consequently not equally 
influenced by the same physical processes, or the time delay for changes that 
affect one region to reach the other is too large for the changes to show up in simultaneous
spectra of both components.

\begin{figure}[htb]
\begin{center}
\epsfig{file=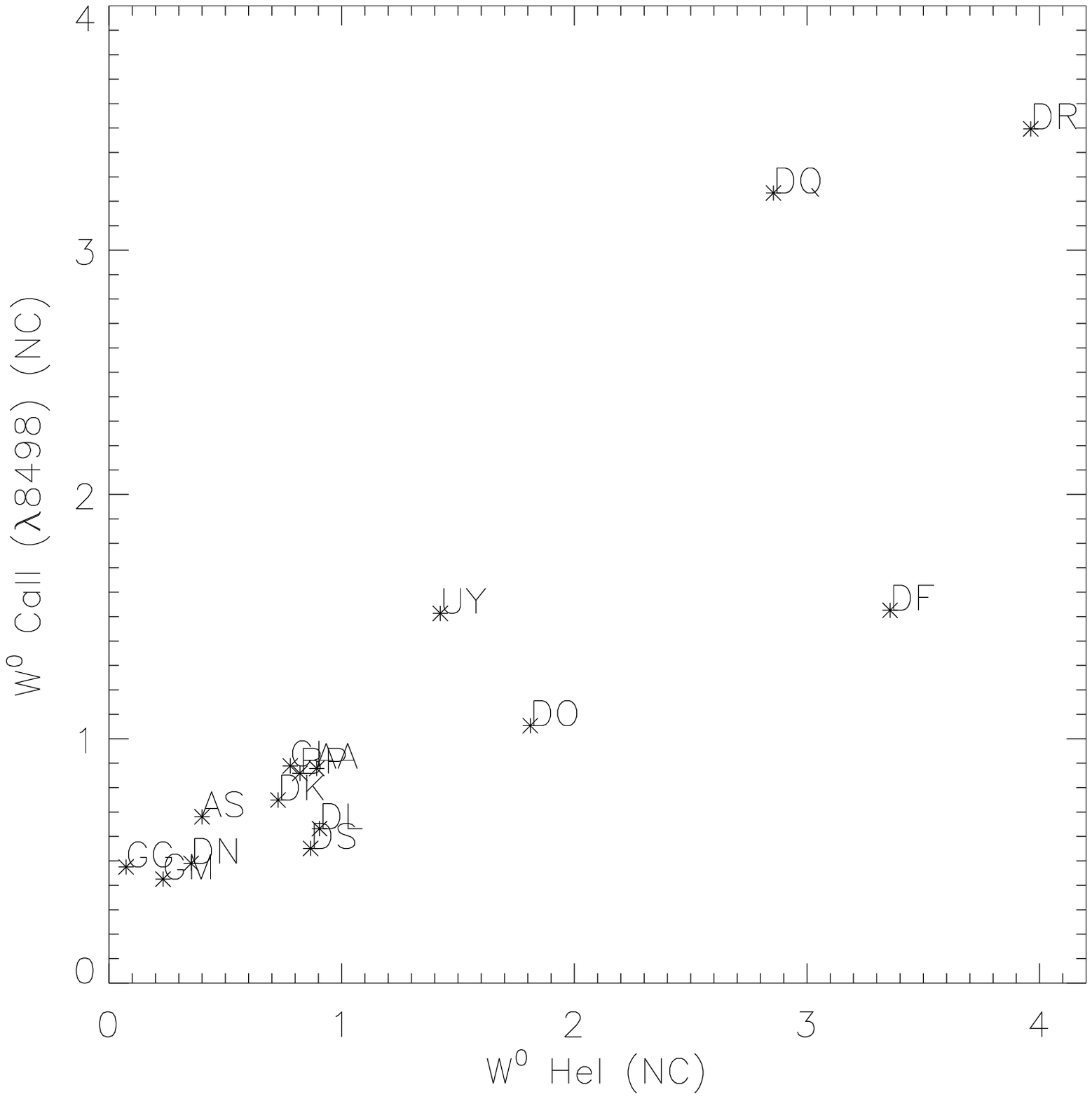,width=8cm}
\caption{\label{wheIcaII} $W^0_{\rm He I}$ (NC) vs. $W^0_{\rm Ca II(\lambda8498)}$ (NC)}
\end{center}
\end{figure}

We also confirmed the trend noticed by \citet{bsb96} that the IRT and the \heI NCs
are positively correlated (Figure \ref{wheIcaII} and FAP=$1.41\times 10^{-5}$), %and r=0.88
(unsurprisingly, since we have a lot of data in common), while \citet{muze} found that they
were uncorrelated. Those lines are generally thought to sample quite different
temperature regions -- the \heI is formed in high temperatures
($> 20,000$K) while the IRT is a chromospheric line formed at $4,000$K $<$ T $< 7,000$K.
The correlation found is probably related to the excess emission due to the 
influence of the accretion shock on the stellar atmosphere, a common feature in both of them.

Iron lines are quite common in CTTSs, so we decided to analyse the behavior of
four of them (2 \feI and 2 \feIIn) in our sample.
The \feII ($\lambda$4923) line was only fitted when in emission. 
Only 40\% of the sample presented this line in emission and we were able to measure the
veiling for just some of them, but before applying the veiling corrections it shows a good 
correlation with the \heI BC and with the IRT BC. 
These correlations may indicate that the \feII lines, at least when in strong emission, 
share a common emitting region with the BCs of \heI and the IRT.

The \feI absorption lines ($\lambda$6192 and $\lambda$5497), with excitation 
energies of 4.43eV and 3.27eV respectively, are not correlated with each other. 
Although visually they both have some resemblance to the \liI line, nothing
was found between the \feI ($\lambda$6192) and \liIn, while a strong correlation exists 
between the \feI ($\lambda$5497) and \liI lines. 
This is due to a temperature effect: the cooler stars show the deepest lines.

\begin{figure}[htb]
\begin{center}
\epsfig{file=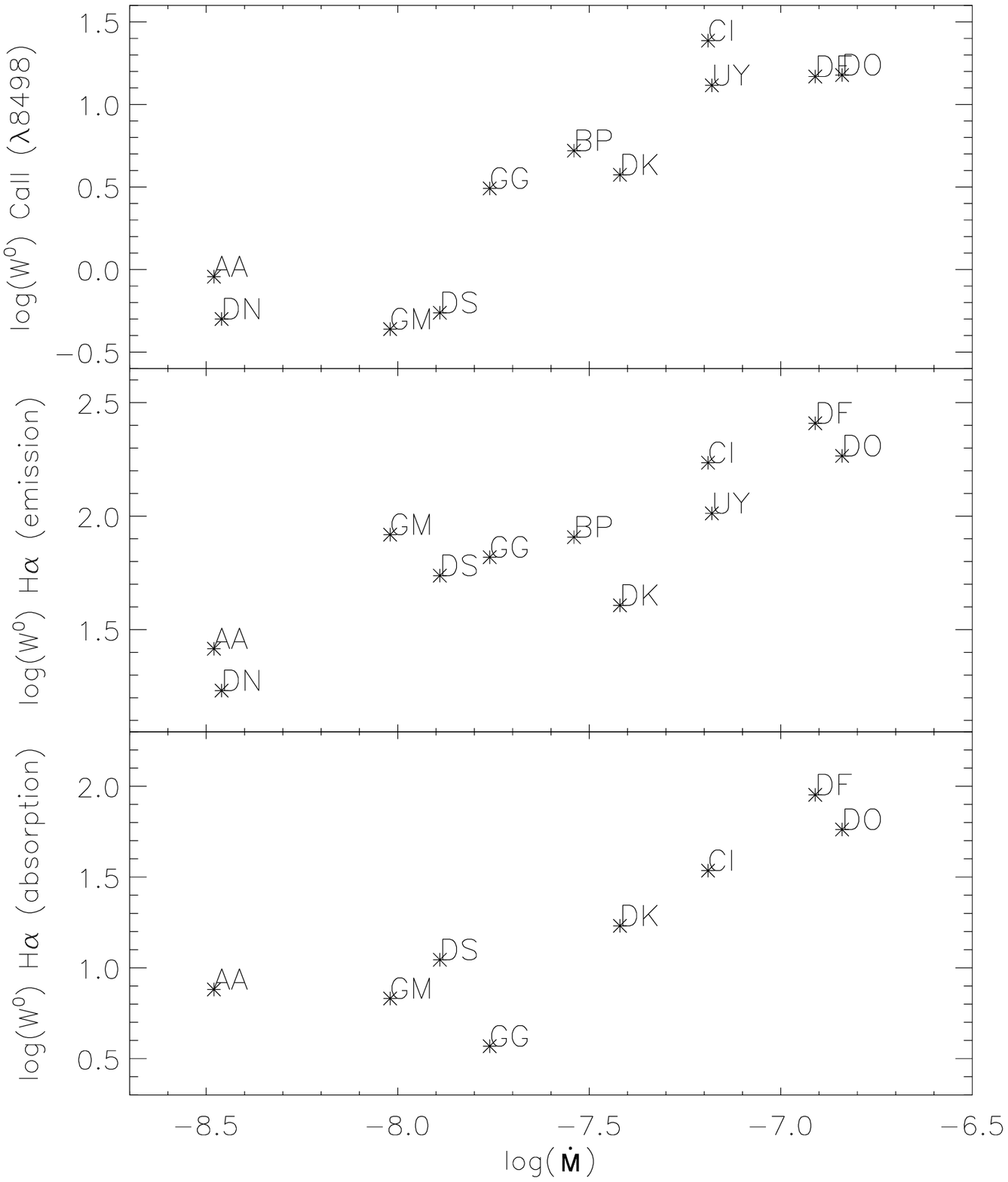,width=8cm}
\caption{\label{macc} $\dot{M}$ vs. $W^0_{\rm Ca II(\lambda8498)}$ and
vs. $W^0_{\rm H\alpha}$ (emission and absorption components). }
\end{center}
\end{figure}

Another important aspect of the correlation measurements is the quest for
good accretion rate indicators that would later provide an easy way to obtain
these rates for other stars. Using the mass accretion rates in \citet{hcgd},
there is a very 
good correlation between the IRT veiling corrected equivalent widths (\caII 
$\lambda$8498 and $\lambda$8662) and mass 
accretion rates (Figure \ref{macc} (top) and FAP=$1.80\times 10^{-4}$), as 
in \citet{muze} for \caII ($\lambda$8542). The only
exception is DQ Tau, that has a $\log{\dot{M}}=-9.4$ \msun ${\rm yr}^{-1}$ 
according to \citet{hcgd} but $-7.3$ \msun ${\rm yr}^{-1}$
according to \citet{heg95} (the largest discrepancies between their values).
DQ Tau is known to have a variable accretion rate that depends on the orbital
phase of the binary system and on events such as outbursts that may occur
when the stars approach each other \citep{bjm}. We decided not 
to use this star.

The Balmer lines and their components are also correlated to the mass accretion rates 
(Figure \ref{macc} (middle and bottom) with FAP=$3.40\times 10^{-4}$ and 
$7.09\times 10^{-3}$ respectively). 
The results for the absorption component are not as reliable as the others, 
mainly due to the small number of stars that exhibit a clear absorption and for which 
there are published mass accretion rates. 
The flux of the emission and absorption components increases as the accretion 
rate increases, showing that the accretion may be powering both the 
emission and the outflow. This would be consistent with theoretical results
\citep{shu94} that suggest that the wind mass loss rate and the disk
accretion rate are directly proportional (but see below for caveats).

\subsection{Evidence for Outflows}

Winds are expected to be present in CTTS and help carry away angular 
momentum. If cool, they should appear mainly as absorption components.
Due to the presence of the disk the absorption is expected
to be blueshifted, the receding part of the wind is blocked by the disk 
from the observer's line of sight. The emission components of CTTS were thought
for some time to be produced in optically thick winds, but these models also
generated upper Balmer line profiles with very deep absorptions that were not
consistent with observations.

The Balmer and Na D blueshifted absorption components are the most 
clear evidence in Figure \ref{profiles} that a strong wind is present in our sample of CTTSs.
Almost 80\% of our stars show blueshifted absorption components in at least
one line, the most common being \hal (see Table \ref{vel}). The number of 
blueshifted components decrease as we look at the upper Balmer lines, 
appearing only in $\sim$ 35\% of the \hdel profiles (where they tend to be weak). 
This suggests that the blueshifted absorptions arise from a region that is optically thin
at least in the upper Balmer lines, and 
distinct from the emission component that is believed to arise mainly in the
optically thick accretion flow. Some stars, like DR Tau and AS 353, present 
P Cygni profiles in all the Balmer lines. These are normally associated with strong winds, 
and other stars, like CO Ori, RW Aur, 
SU Aur and RY Tau, have such a strong absorption component that it sometimes 
suppresses the emission in the upper Balmer lines of the stars 
with the hottest underlying photospheres. A number of these stars are known to be jet
sources. 

\begin{figure}[ht]
\begin{center}
\epsfig{file=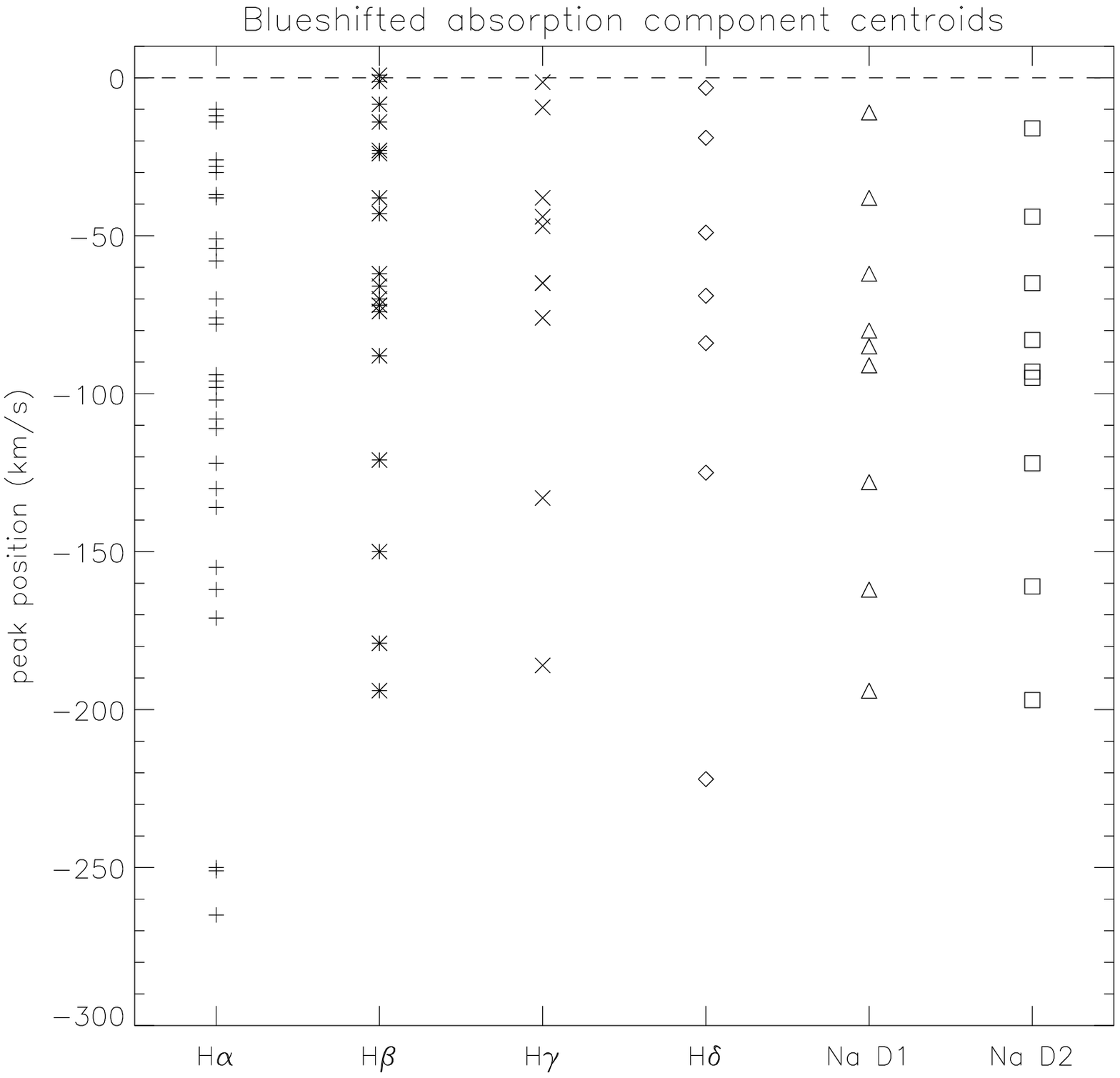,width=8cm}
\caption{\label{blue} Blueshifted absorption component's peak position.}
\end{center}
\end{figure}

The 8 stars in our sample that exhibit blueshifted absorption components in the
Na D lines \citep[presumably indicating high mass loss rates;][]{hart98},
also do in \haln. The lack of P Cygni profile in the Na D lines
does not necessarily mean low mass loss rates, as the star may present a deep 
photospheric absorption that prevents the detection of the blue absorption due to 
the wind \citep{natta90}.
Due to the difference in their optical depths,
the Balmer and Na lines will sample different wind volumes. As resonance 
lines, the sodium probes the coolest part of the wind. 
The blueshifted components of both lines tend to be centered at approximately 
the same velocities (Table \ref{vel}), likely meaning that different wind regions 
share some common kinematic component. 
The Na D blueshifted absorptions bear no other resemblance
to the \hal ones, being much weaker in intensity and width.
The position of blueshifted absorption centers in these lines vary from star 
to star suggesting that the wind velocities can occur with a wide range of values 
(from 0 to --270 \kms in our sample, see Figure \ref{blue}). There may well be 
projection effects that must be taken into account; these would imply
the outflow is not spherical.

After decomposing the profiles we can also investigate the correlations that 
appear between different wind components
and between wind components and other features in order to better understand the 
wind forming region.
The \hal and \hbeta blueshifted absorptions are very well correlated 
(FAP=$2.39\times 10^{-8}$), % and r=0.986 
showing that they both probe the same wind region, and
although there is apparently no correlation with the blueshifted absorptions
of \hgam and \hdeln, this may be due to the small number of cases. Where the 
absorption does appear in the upper lines the wind component 
is often very weak and consequently hard to fit.

The \hal and \hbeta blueshifted absorptions also show some correlation with
their line's emission components: the stronger the emission, the stronger the 
absorption. Unfortunately, that does not have a unique interpretation.
One possibility is that the same mechanism that is powering the 
emission is also powering the wind in the sense that when more material is 
accreted there is more to outflow. This is predicted by the accretion model
and also means that the absorption region is outside the emission region. A 
second possibility is that if
both emission and absorption were formed in winds, the absorption component 
would be due only to the decay of the source function in the outer regions of
the atmosphere and not to the accretion itself. If the first possibility is
correct, this relation could also explain 
the trend seen between the \hal emission and absorption components and the mass 
accretion rates.

Finally, the present result may also be an effect of line optical depth;
if the wind is optically thick the central depth of the blueshifted
absorption is fixed and the changes that appear are in fact only due to
the emission component variations as shown by \citet{jb95b} in the case of SU Aur.
\citet{jb97} also found that the \hal absorption and emission strengths of many
DF Tau spectra were correlated and that the absorption component strength correlated with
the veiling.

\begin{figure}[ht]
\begin{center}
\epsfig{file=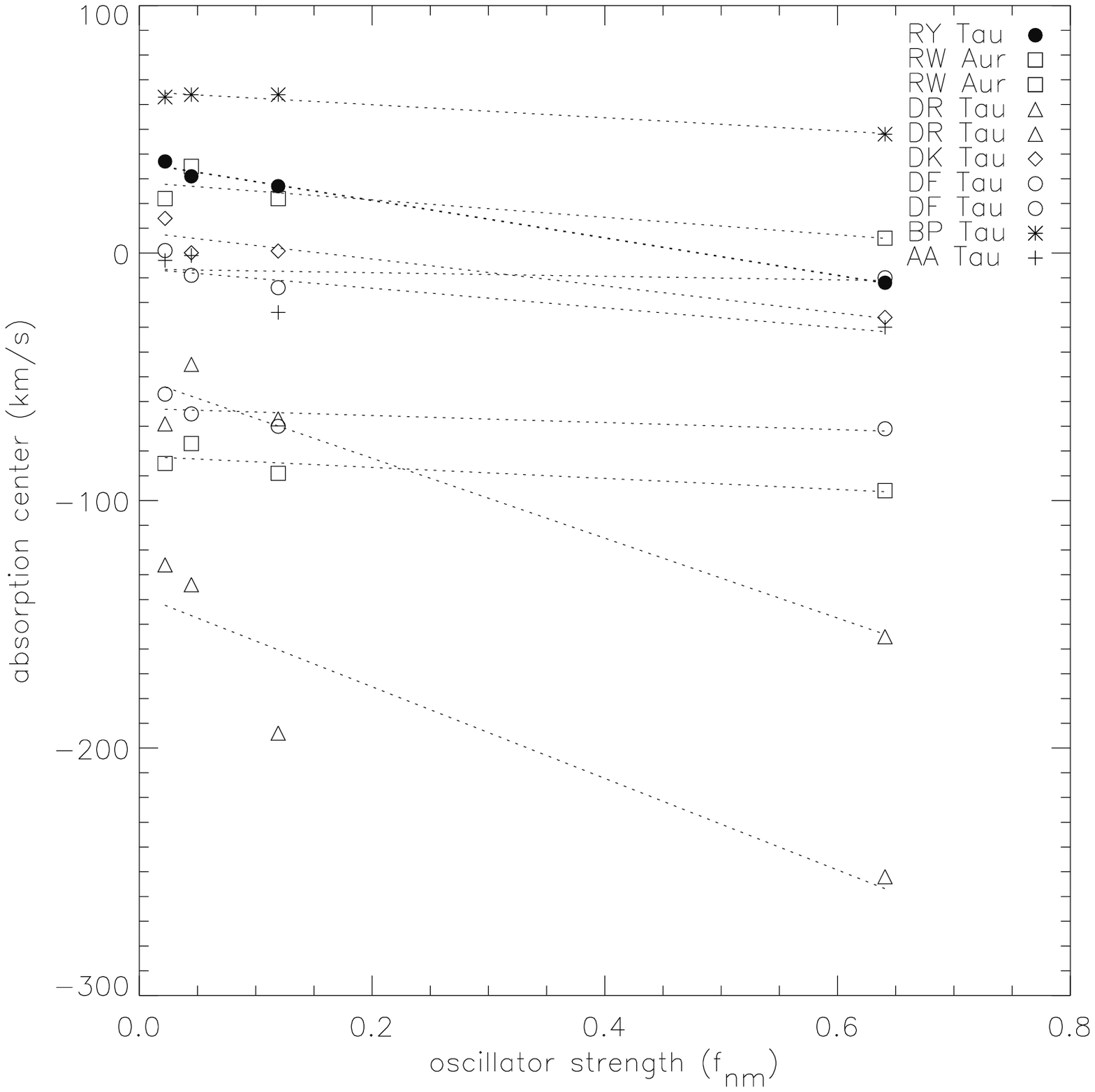,width=8cm}
\caption{\label{gradient} Balmer line absorption center vs. oscillator strength.
RW Aur, DR Tau and DF Tau are plotted twice as 2 wind components were measured.}
\end{center}
\end{figure}

A curious fact about the absorption components can be noticed in the 
stars that show an absorption component in the 4 analysed Balmer lines.
This component tends to be shifted towards the red 
as we go from \hal to \hdel and in Figure \ref{gradient} we plot the 
absorption centers vs. the Balmer lines oscillator strengths for those stars. 
It seems that we may be seeing the acceleration of the flow. 
This shift of
the absorption component can also be noticed in \citet{ehg94}
in the residual profiles of DK Tau and in \citet{appenz} 
in the DR Tau spectra, showing that this behavior may be quite common
among CTTSs. 

Another feature that is associated with
strong activity, presumably both inflow and outflow, 
are the iron lines in strong emission.
These lines are often in absorption in our sample but
\feII ($\lambda$4923) is also found in emission in $\sim$ 40\% of our stars.
When this happens, a wind component is present at least in \haln,
sometimes also in the other Balmer lines and in the Na D lines, and the
IRT lines always show a BC. 
The Fe lines in emission seem to be related to both the presence of
outflow and high veiling, as some stars that show blueshifted absorption
in \hal but no BC in the IRT do not present \feII ($\lambda$4923) in emission.
DG Tau, AS 353 and RW Aur, that also exhibit \feI ($\lambda$6192) and \feII ($\lambda$4352)
in strong emission, also have associated jets \citep{ehg94,mundt}.
But unlike the forbidden lines that tend to be blueshifted, the iron lines when
in emission are centered at the stellar rest frame. They are probably not 
produced in the outflow (wind or jet), since the occultation of the receding
part of the flow by the disk would yield a blueshifted line. 

\subsection{Evidence for Infall}

Redshifted absorption components in the line profiles
(inverse P Cygni profiles), can be the clearest
spectral evidence that infall of material is occurring, and are naturally predicted
by the magnetospheric accretion models.
These models, however, require special conditions for the redshifted absorption
component to be visible, such as low line thermalisation,
low emission damping wings, and proper inclination values. The inclination
dependence arises from the contrast of the line source function of the infalling
gas and the continuum source function where the gas stream is projected
\citep{hart94}. If the projection is against the cool photosphere (low
inclinations) no absorption is found, but if it is against the hot shock
material, like a ring or spots, (higher inclinations) the absorption can be
produced.

\citet{ehg94} presented an analysis of 15 CTTSs
included in our sample, and showed that 87\% of their stars had redshifted
absorption components in at least one line at $\sim$ 200 to 300 \kmsn,
a range of velocities that
is consistent with the ballistic infall predicted by the magnetospheric models
for a typical CTTS. Only $\sim$ 40\% of our stars have
redshifted absorption components in at least one line and generally not 
at $\sim$ 200 to 300 \kmsn. BP Tau, for example, shows redshifted absorption in many
lines at $\sim$ 50 \kms and UX Tau has a strong redshifted absorption in \hal
at 43 \kmsn. The difference in the absorption centroid position of the
stars from the values predicted by the models could partially be due to inclination effects.
The difference in the number of redshifted absorption
occurrences found in the two different analysis could be due to the fact that
\citet{ehg94} analysed residual profiles instead of the original ones. This
gives, according to them, a high sensitivity in defining features such as
the Balmer lines wings, where the redshifted absorption lies. However,
in the residual profiles showed by \citet{muze} of 11 CTTSs, redshifted
absorption at typical free-fall velocities can be found in less than
half of their sample, which is more in agreement with the predictions of
the magnetospheric accretion models.

In order to compare our results with the ones cited above
we decided to generate residual profiles from our sample, choosing 8 lines
commonly found in emission:
\haln, NaD, \heIn, CaII ($\lambda$8498 and $\lambda$8662), \hbetan, \hgam and
FeII ($\lambda$4352). \hdel was excluded as most of our spectra present low
signal to noise in that region, making it hard to obtain reliable information
about small absorption features. We generated the residual profiles by subtracting
from each observation a broadened and veiled standard spectrum, using
as standard stars the ones previously selected for the veiling calculations.

\begin{figure}[ht]
\begin{center}
\epsfig{file=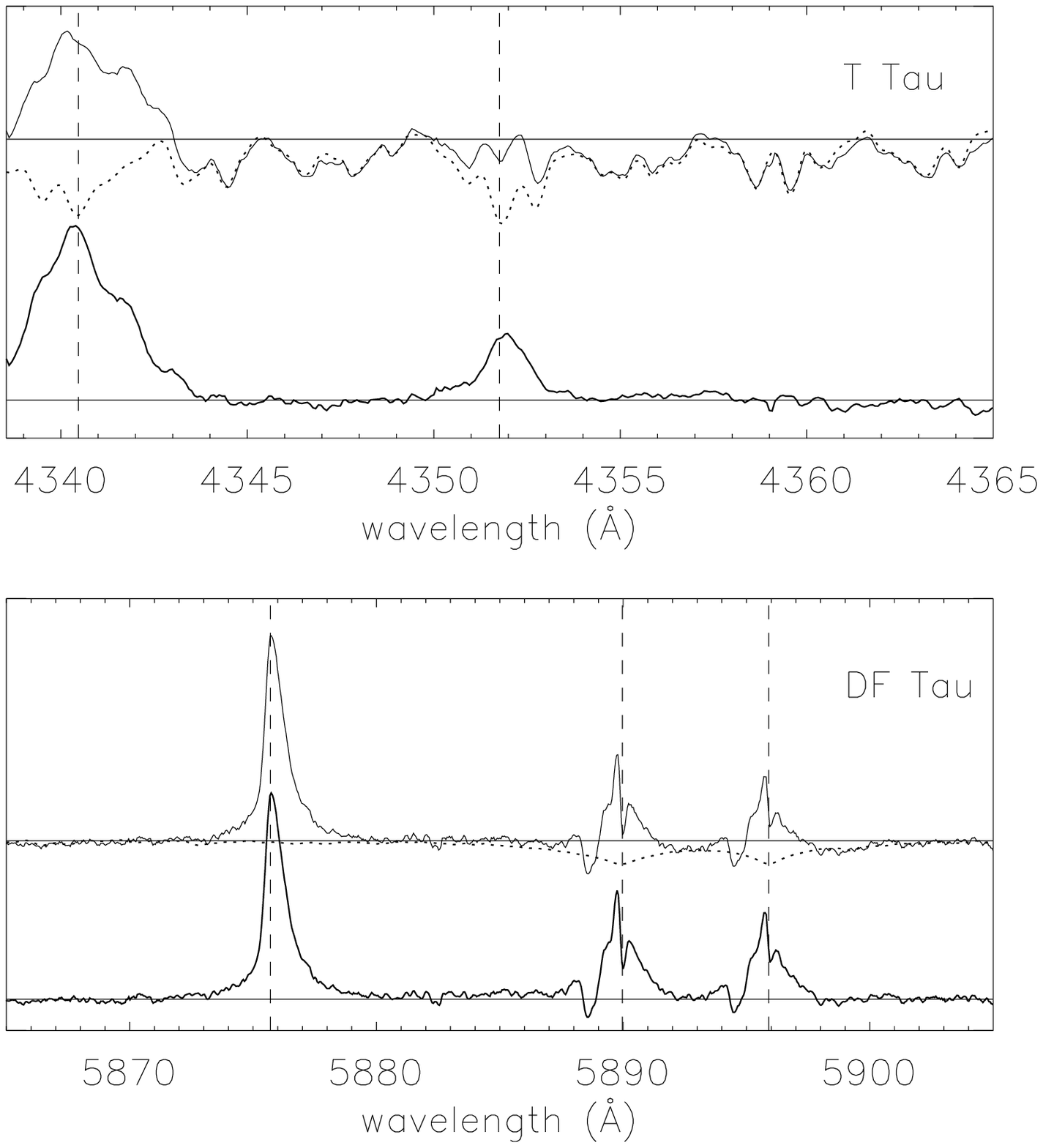,width=8cm}
\caption{\label{residual} Residual profile determination, the thin solid line
corresponds to the
original spectrum, the dotted line to the standard spectrum and the thick solid
line to the residual CTTS profile. Top: \feII ($\lambda$4352) is dramatically
enhanced by the residual method. Bottom: the redshifted absorption that appears
in the unsubtracted spectrum in the Na D lines turns out to be mainly photospheric.}
\end{center}
\end{figure}

Although the residual method often enhances features like weak emission
lines (see Figure \ref{residual}, top), care must be taken during the subtraction, 
as a small difference in continuum
normalization between the two spectra can also introduce false features.
The uncertainty in the veiling determination and the difficulty of choosing
suitable standard stars 
are other factors that can introduce errors in the process.
Finally, the legitimacy of the concept itself is dubious when the overlying 
emitting region is optically thick over most of the profile.

The residual profile calculation suppresses, in principle, all the photospheric
contribution that is present in the spectra,
generating smoother profiles, as the broad lines are no longer superimposed with
narrow absorption ones. However, we noticed that it did not a priori enhance
the redshifted absorptions present in our unsubtracted spectra; indeed some of the 
redshifted absorption components turned out to be mainly photospheric 
(for example, the Na D lines of BP Tau, DF Tau, DQ Tau and RY Tau, see 
Figure \ref{residual}, bottom).
In general, most of the redshifted absorptions 
became shallower, although less noisy, and we did not find many new ones. Our 
residual profiles show that $\sim$ 50\% of our stars present at least one line with 
a redshifted absorption component, a result in agreement with the original 
(unsubtracted) data, with the magnetospheric models, with the results by 
\citet{muze} but not with \citet{ehg94}.
Although very weak features such as the iron emission lines and the redshifted
absorptions at free-fall velocities are strongly affected by the underlying
photospheric spectrum, one may notice that the general characteristics of the 
line profiles discussed in this work are weakly affected by the subtraction of 
the photospheric features and that our results are also valid for residual 
profiles.

Finally, in agreement with both the model predictions and with \citet{ehg94} we
see less evidence of redshifted absorption in \hal at high velocities than 
in the upper Balmer lines, probably due to thermalisation effects and the Stark 
damping of the \hal wings.

\subsection{Profile Symmetry}

The magnetospheric accretion models predict that the BC emission is formed
in the magnetospheric infalling material. It should then exhibit a
central or slightly blueshifted peak and blueward asymmetry due to occultation
by the disk and star of part of the flow moving slowly away from the observer.
By blueward asymmetry we mean that there is an excess of blue emission compared
to the red emission.
Some theoretical profiles published by \citet{hart94} also show redshifted absorption 
at free-fall velocities, while others show extended red wings caused by the 
occultation of the high velocity blueshifted material by the star; both effects 
add up to enhance the line asymmetry.

One way to search for asymmetries is to reflect one of the line wings about 
the broad emission centroid, as done by \citet{jb97} for DF Tau. 
If the wings match each other, the line is considered symmetric, and
this avoids a confusion between lines that are only shifted and lines that
are really asymmetric. P Cygni profiles are more difficult
to evaluate, since the blue emission wing is sometimes completely absent and it may be
hard to determine the correct shift to be applied.

\begin{figure}[ht]
\begin{center}
\epsfig{file=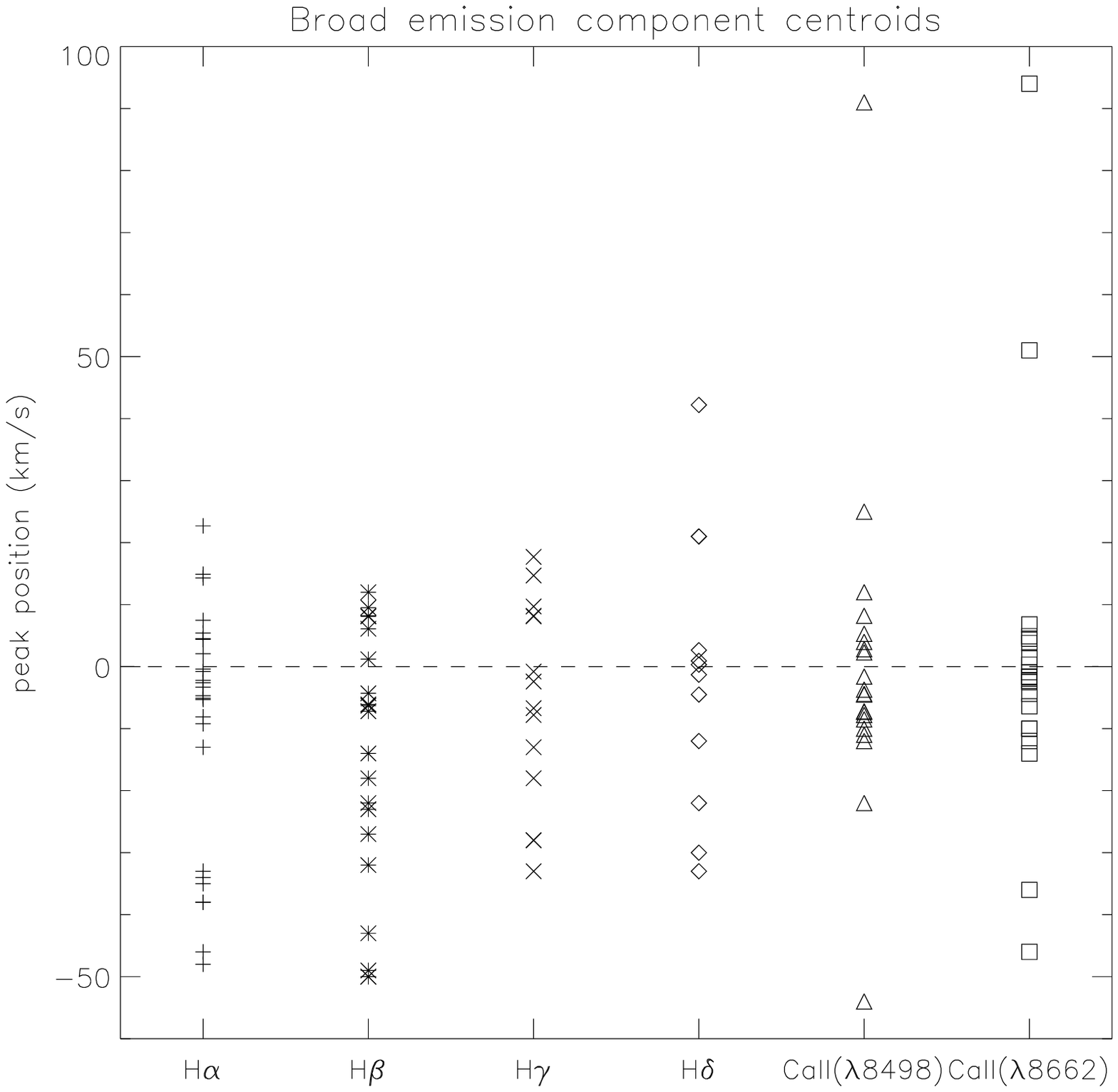,width=8cm}
\caption{\label{peak} Broad emission component's peak position. Lines with P Cygni
profiles and those not reliably measured are not included.}
\end{center}
\end{figure}

In Table \ref{reflection} we show the velocities that produced the best agreement 
between the opposite wings, and we also indicate which parts of the line are symmetric 
and which are not in the case of partially symmetric profiles.
The BCs that showed blueward asymmetry are labelled ``ASb'' in Table \ref{reflection},
but they comprise only $\sim 20\%$ of the analysed lines. Most of the BCs are
symmetric (well fitted by a Gaussian) and $\sim 75 \%$ of them have a central or
slightly blueshifted peak. 
In Figure \ref{peak} we show the velocity position of the
broad emission peak for several lines and we notice that, although most
of them are indeed blueshifted, a substantial number present a redshifted peak.
This is not seen in the published theoretical magnetosphere profiles.
Lines with P Cygni profiles and those where the emission component was not 
reliably extracted are not included in this plot.

\begin{figure}[ht]
\begin{center}
\epsfig{file=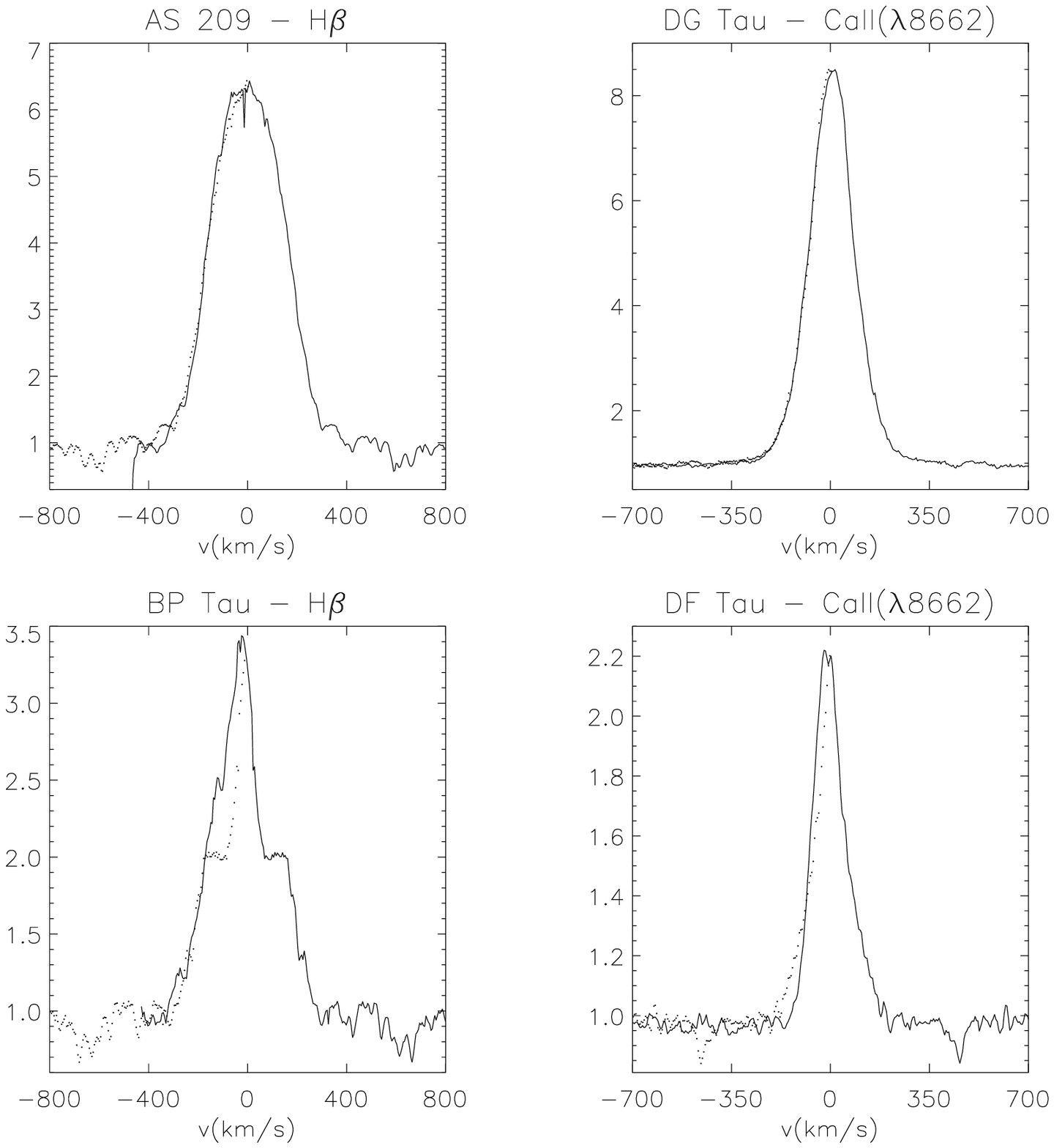,width=8cm}
\caption{\label{symmetry} Emission lines with the red wings (dotted lines) reflected.
The top lines are
symmetric and the bottom ones are asymmetric and in agreement with the magnetospheric
model predictions.}
\end{center}
\end{figure}

Among the lines that presented asymmetries, the ones identified as ``${\rm ASb}^b$''
have symmetric low velocity material and a blueward excess of high velocity emission,
and those identified as ``ASr'' show redward asymmetries. 
Neither type is predicted by the magnetospheric model.
Some lines do present profiles with the characteristics described
by the magnetospheric models, corresponding to ``ASb'' and ``${\rm ASb}^c$'' 
in Table \ref{reflection}.
Examples of profiles in agreement and opposite to the
magnetospheric symmetry predictions are shown in Figure \ref{symmetry} with the 
reflected wings overplotted.

Each \caII IRT line is blended with a Paschen emission line
and this could lead to a wrong asymmetry classification. However, 
from the thirteen stars that presented IRT asymmetries only
four could be due to the Paschen contribution, which affects 
the high velocity wing redward of the line center. These stars are
quoted as uncertain asymmetry measurements in Table \ref{reflection}.

In general the symmetry predictions of the magnetospheric model cannot be
strongly confirmed with our sample. Contrary to them, most of the BCs are
found to be symmetric; among the asymmetric ones many are not blueward asymmetric,
and $\sim$ 20\% of the broad emission components present centroids redshifted by
more than 5\kms.

\section{Discussion}

Many aspects of the line formation of a CTTS can be inferred from the correlations 
found with the veiling corrected equivalent widths in the previous sections. 

We showed that lines coming 
from a common lower level are very well correlated, like \hbeta with \hal and the 
IRT lines with each other. This is expected, as they need the same physical 
conditions to be produced and tend to be affected by the same physical processes. 
As previous studies also showed,
most lines have more than one component (BC and NC for the \heI and
the \caII and emission and absorption components for the Balmer and Na D lines).
These are likely formed in distinct regions, as the characteristics of those components 
are very different from each other throughout our sample and many correlations 
found were valid for one of the components but not for the other. 
According to our data %a general picture of the line formation in CTTSs would be that 
the Balmer emission, the \heI BC, the IRT BC and the iron lines 
(when in strong emission) are mainly formed in the magnetospheric flow. Alternatively,
the \heI BC may also come from the shock region, where the temperatures are
high and more appropriate for the formation of that line instead of
the magnetospheric flow itself. The NCs of the Ca and \heI lines are thought to be formed 
at the stellar surface where it is perturbed by the infall of material and
partly in the normal stellar chromosphere. The wind is the origin of
the blueshifted absorption commonly seen in the Na D and Balmer lines.

We also verified that
the effect of accretion on the line components sometimes probably produces
the correlations, as lines formed in regions with physically distinct properties 
(very different temperatures, for example) but both affected by accretion can be found 
to correlate very well, like the \heI and the IRT NCs.

Somewhat puzzling is the lack of strong correlation between the IRT BCs and the \hal
emission or the \heI BCs, if all broad emission lines are supposed to share a 
common forming region.
\citet{jb97}, analysing DF Tau, confirm the lack of correlation found between the IRT BC 
and the \hal emission but they found a correlation between the \heI BC and the
IRT that neither we nor \citet{muze} confirm in our samples of various CTTSs. 

We showed that the IRT lines and \hal are good mass accretion indicators in our sample.
We also confirmed that the same mechanism that powers the accretion also
seems to influence the outflow, as the emission and absorption components of \hal are well 
correlated with the mass accretion rates and with each other. Other good evidence
for the accretion-wind connection is that the Fe lines in strong emission seem to be
related to both processes, occurring when the wind and the veiling are prominent.

The blueshifted absorption components are thought to be formed in low density winds. 
There is strong evidence that the NCs are formed at the stellar surface
\citep{bsb96}. Although they are both influenced by the accretion process, they 
should not be compared to the theoretical magnetospheric profiles. 

Some of the observations tend to support the fact that 
the accretion affects the relatively quiet stellar atmosphere, and lines produced in
that region suffer the influence of such changes. The models also predict that accretion
and outflow should be related to each other -- the former occurring through closed
magnetic field lines that connect the disk to the star and the latter through the
open field lines escaping from the disk itself. 
 
However, the results presented in the previous sections also show that some of
the general characteristics of the magnetospheric accretion models are not always 
confirmed by our data. The idea that the broad emission components mainly arise 
from a common emitting region, the infalling magnetospheric gas according to 
those models \citep{hart94}, must be taken with care, as the observational 
evidence is not yet conclusive.
We must also keep in mind that the magnetospheric predictions do not do 
particularly well for the broad emission components that they primarily apply to.
We find observed profiles to be generally fairly 
symmetric with centroids distributed through a range of blueshifted to redshifted velocities. 
Sometimes they have redshifted 
absorption components, but often not at free-fall velocities. Only 
about 20\% of the BC profiles support the specific predictions of the models.

\citet{hart94}, \citet{muzeb} and \citet{muze} presented line profiles
generated with magnetospheric models and compared them with observational 
ones. They obtained good agreement between the observed and theoretical profiles
of many lines for some stars, like BP Tau. But comparing their BP Tau 
observed profiles to ours, many differences can be seen. This is expected, 
as CTTSs are known
to exhibit line profile variability sometimes even within a few hours \citep{jb95}, 
and \citet{gull} showed that this is the case for BP Tau, a typical CTTS.   
The problem that arises with the rapid profile variation is that, even if the
magnetospheric models can reproduce one of the profile types of a star,
it does not mean that all the profiles that a line may exhibit will be
reproduced without changing basic model characteristics. As an example,
the model used to compare the observed and theoretical profiles of BP Tau cited above
does not fit our observations of that star well.

Some years ago, \citet{shu94} proposed a new version of their X-celerator mechanism
whose natural period was the rotational period of the star. Lines formed
in magnetospheric accretion flows that are controlled by the stellar field might
also show periodicity at the stellar rotation period in the presence of an
inclined dipole field. In this case, the line variations
would certainly be due to geometrical effects and a model with the same basic 
parameters should be able to describe the various line profiles. SU Aur presented this
kind of variations in \hbeta and \hal \citep{jb95b}, but unfortunately its profiles 
do not resemble the published theoretical ones very much. 

In general, however, the line 
variations do not seem to be related to orbital motion as they do not usually 
correlate with the stellar rotational period. Different parts of a line may vary 
differently as \citet{jb95} showed by analysing time variations of \hal for several CTTSs. 
They suggest the line emission region is composed of discrete, stochastically
varying blobs with a range of velocities and turbulence, which the Sobolev treatment
used in the actual magnetospheric models does not take into account.

\citet{najita}, analysing Br$\gamma$ emission profiles, showed that 
WL 16, an obscured
low-luminosity YSO, presented profiles that were remarkably similar to the theoretical
Balmer line calculations by \citet{hart94}.
Later, \citet{muzeb} calculated profiles for the Br$\gamma$ emission and 
showed that, in fact, the WL 16 Br$\gamma$ emission line was very well reproduced. However,
some of the stars in the sample of \citet{najita} also present very symmetric emission
profiles, like AS 353, DG Tau and SVS 13, that cannot be explained only
by magnetospheric accretion.
Although the magnetospheric model can generically explain part of the large 
line widths and the 
occasional redshifted line absorptions, it does not predict symmetric profiles
because the magnetospheric infalling material is subjected to occultation effects by the
star-disk system.

Most of our broad emission profiles were found to be symmetric and the few
asymmetries occur on both the blue and the red side. Due to these 
differences, the 
profiles presented in the theoretical papers generally do not match our observed ones.
Many authors have speculated on the origin of the rather symmetric broad line wings in
TTS profiles. \citet{basri90} suggested that the wings were formed in a turbulent
region near the star and \citet{ehg94} pointed out that Alfv\'en waves could generate
the necessary turbulent broadening. \citet{jb95b} managed to reproduce some of 
the Balmer line symmetric features of SU Aur by adding a range of turbulent 
high velocity components 
at the base of spherically symmetric wind models and \citet{jb95} could also fit 
the DF Tau wings with a similar procedure.

Most of the stars discussed here rotate slowly ($v\sim 10$\kmsn)
at velocities that are much smaller than the velocities of the infalling 
material ($v\sim 200-300$\kmsn) and the inclusion of rotation in the 
theoretical models will not significantly alter the profiles in this case.
Even in the case of a stiff magnetosphere, material coming from a disk
at $3R\star$ would have $v\sim 30$\kms and would not influence the entire line.
The addition of winds to the theoretical models will certainly
change the line shapes but stars like AS 209 and DE Tau, that do not always
exhibit blueshifted absorption and are rather slow rotators, will still need
some further explanation of their broad symmetric line profiles
(Figure \ref{profiles}).

Stark broadening may explain the low very broad far wings in \hal
\citep{hart94,muzeb}, but will not produce the main Gaussian shape commonly
present in many different lines. A better treatment of the radiative transfer
(actually done with the Sobolev method) could also alter the theoretical line profiles
and would properly take into account Stark and opacity broadening effects.

Magnetospheric accretion probably does occur in CTTS. Evidence for it includes
hot spots at the stellar surface of these stars, thought to be due to the 
magnetic accretion shock, redshifted absorption 
components at free-fall velocities that would be hard to explain without 
magnetospheric infall, as well as cases like SU Aur and WL 16 that give more explicit
support to the theory. However, it may not be the only important mechanism. 
The symmetry of the 
line profiles and the lack of correlation of the variation among different line 
regions suggest an important turbulent process may be occurring together with 
the magnetospheric infall.

\section{Conclusions}

We have presented the spectral analysis of a sample of CTTSs covering a wide
range of optical wavelengths and veilings.  
We tested the predictions of the magnetospheric accretion model and previously published
results supporting them. We confirm that
many CTTSs exhibit central or blueshifted broad emission lines but a substantial minority
show redshifted centroids. 
The accretion and outflow seem to be related and redshifted absorption 
components are sometimes present in the spectra. We found that the analysis 
and interpretation of these is not straightforward. We could not reproduce the 
observational results that showed 
a very high frequency of redshifted absorption in the lines of CTTSs.
We also show that most of our emission components are symmetric,
instead of blueward asymmetric as predicted by the theory 
or suggested by previous observational studies. We saw that the broad components of different
atomic lines are not strongly correlated to each other, while we would expect them
to be if they were all formed in the accretion flow.

Our results do not refute the general magnetospheric
accretion scenario, but rather indicate that is only part of the important processes
which produce the strong permitted line emission in CTTSs. Our intent is
to point out that the case has not yet been strongly made, and
that much more observational and theoretical work is required. Part of the emission line 
profiles may also be produced in winds, and rotation and turbulence must be added to 
the models.

Synoptic observations are critical to understanding the full range of profiles produced
by a single star, where some underlying parameters do not change but others do,
and many structures are dynamic.

\acknowledgments
This research is based on data collected on the Shane 3m telescope at
Lick Observatory run by the University of California. We would like to
thank Christopher Johns-Krull, Anthony Misch, Michael Bisset, Claude Bertout,
Natalie Stout-Batalha and Celso Batalha who helped gather the profiles presented here.
S.H.P.A. acknowledges support from the Conselho Nacional de Desenvolvimento 
Cient\' {\i}fico e Tecnol\'ogico - Brazil.

\newpage

%--------------------------BIBLIOGRAPHY---------------------------

\newpage

\begin{table}
\dummytable\label{eqw}
\end{table}

\begin{table}
\dummytable\label{vel}
\end{table}

\begin{table}
\dummytable\label{reflection}
\end{table}

\end{document}

\newcommand \caIIn{\ion{Ca}{2}}
\newcommand \haln{H$\alpha$}
\newcommand \hbetan{H$\beta$}
\newcommand \hgamn{H$\gamma$}
\newcommand \hdeln{H$\delta$}
\newcommand \kmsn{km s$^{-1}$}
\newcommand \nosp{$\!\!\!\!$}

\begin{deluxetable}{lcccccc}
%\scriptsize
\tablewidth{0pt}
\tablecaption{\label{reflection} Reflection velocity (\kmsn) to match the 
red and blue wings.}
\tablehead{
\colhead{Star} & \colhead{\haln} & \colhead{\hbetan} 
& \colhead{\hgamn} & \colhead{\hdeln} & \colhead{\caIIn} & \colhead{\caIIn} \\ 
\colhead{} & \colhead{} & \colhead{} & \colhead{} & \colhead{} 
& \colhead{\scriptsize $\lambda$8662}& \colhead{\scriptsize $\lambda$8498}}

\startdata
AA Tau & -30    & ASb    &  ASb:  &  ASb  & $0^a$ & $-14^a$ \\
AS 209 &   2    &  2     &    10  &    0  &   5   &   -7     \\ 
AS 205 & $0^c$  &\nodata &\nodata &\nodata&  -10  &  -10:\\ 
AS 353 &   0    &  20:   &    0:  &  0:   &   0   &  -10 \\ 
BP Tau & ${\rm ASb}^c$ & ${\rm ASb}^c$ & ${\rm ASb}^c$ & ASb & 0 & 0 \\ 
CI Tau & -10    &\nodata &\nodata &\nodata&   0   &  -5 \\ 
CO Ori &  23    &  x     &   x    &   x   &  x    &  x  \\ 
CW Tau &   0    &  10    &   x    &   x   &   5   &  ${\rm ASr}^b:$ \\ 
DE Tau &   0    &  12    & ${\rm ASb}^c$ &\nodata  &   7:  &  0    \\ 
DF Tau &   0    & -18    &  -27   &  ASb: & ASb: & ASb:\\ 
DG Tau & -33    & -33    &   20   &   20  &  5    &  ${\rm ASr}^d$ \\ 
DK Tau & ${\rm ASr}^c$ &  ASb:  &   ASb  &  ASb  & ASr:  & ASr: \\ 
DL Tau &   0    &  8     &   10   &   20  &   0   & ${\rm ASr}^d$   \\ 
DN Tau & ${\rm AS}^{\ast}$ & ${\rm ASb}^b$   & 0   &    0  & $0^a$ & $-8^a$ \\ 
DO Tau &   0    &  -3    &   x    &   x   &  -6   &  -4    \\ 
DQ Tau & ${\rm ASr}^d$ &${\rm ASr}^d$&  -6    &  -10  & ASb   & ASb \\ 
DR Tau &   0    &   0    &  -20   &    0  & ${\rm ASr}^b:$ & ${\rm ASr}^b:$ \\ 
DS Tau & -35    & -50    &  -30   &  -35  & $-9^a$& $-12^a$\\ 
GG Tau & ${\rm ASb}^c$: &   0    &    5   &   40  & ASr   & 0 \\ 
GK Tau & ${\rm ASb}^c$& -35    &  -25   & \nodata  & $-4^a$:& $-5^a$\\ 
GM Aur & ${\rm ASb}^d$ & -50:   &  -30   &  ASb  & $0^a$ & $-5^a$\\ 
GW Ori &   0    & -22    &   x    &   x   & ${\rm ASb}^b$: &   0 \\ 
HL Tau &   0    & \nodata &\nodata&\nodata& ${\rm ASb}^c$&  -8 \\ 
RW Aur & ${\rm ASr}^b$: &  10    &   x    &   x   & ${\rm ASb}^c$& ASb \\ 
RY Tau & ${\rm ASr}^d$&  x     &   x    &   x   & ASr   & 5   \\ 
SU Aur & -13:   &  x     &   x    &   x   &  x    &  x  \\ 
T  Tau &   0    &  -5    &   15:  &    0  &  -5   & -10 \\ 
UX Tau &  -42   &\nodata &\nodata &\nodata& $-8^a$& $-15^a$\\ 
UY Aur & ${\rm ASb}^c$  & -15    &  -20   & \nodata  & ASb:   & ASb: \\ 
V410 Tau & ASb & x & x & x & x & x \\ 
\tablecomments{Lines identified as ``\nosp\nodata\nosp'' were not available and as ``x'' 
were not reliably measured or observed in absorption. ASb means that the line
is blueward asymmetric and ASr redward asymmetric. $^{a}$=only the narrow 
component is present, $^b$=only the low velocity part is symmetric, the wings 
at high velocities are asymmetric,
$^c$=only the high velocity is symmetric, the rest of the line is asymmetric,
$^d$=the high and low velocities are asymmetric, the line part between those is
symmetric.}
\enddata
\end{deluxetable}